\documentclass[prl,aps,twocolumn,reprint,floatfix,nofootinbib]{revtex4-1}
\usepackage{graphicx,epsfig,amssymb,amsmath}
\usepackage[usenames]{color}
\usepackage[normalem]{ulem} 
\usepackage{natbib}

\newcommand{\beqn}{\begin{eqnarray}}
\newcommand{\eeqn}{\end{eqnarray}}
\newcommand{\beq}{\begin{equation}}
\newcommand{\eeq}{\end{equation}}

\newcommand{\ta}{\alpha}
\newcommand{\tk}{\mathbf{k}}

\usepackage{graphicx,epsfig,amssymb,amsmath}
\usepackage[usenames]{color}
\usepackage[normalem]{ulem} 

\usepackage{natbib}
\usepackage{graphicx}
\usepackage{subfigure}
\usepackage{amssymb}
\usepackage{amsmath}
\usepackage{mathrsfs}
\usepackage{multirow}
\usepackage{array}
\usepackage{threeparttable}
\usepackage{color}
\usepackage{hyperref}
\hypersetup{
    bookmarks=true,
    bookmarksopen=false,
    colorlinks=true,        
    linkcolor=blue,          
    citecolor=blue,        
    filecolor=magenta,      
    urlcolor=cyan           
}

\begin{document}

\title{Spontaneous decay of periodic magnetostatic equilibria} 
\author{William E.\ East, Jonathan Zrake, Yajie Yuan, and Roger D.\ Blandford}
\affiliation{ Kavli Institute for Particle Astrophysics and Cosmology, Stanford
  University, SLAC National Accelerator Laboratory, Menlo Park, California
  94025, USA }

\begin{abstract}
  In order to understand the conditions which lead a highly magnetized,
  relativistic plasma to become unstable, and in such cases how the plasma
  evolves, we study a prototypical class of magnetostatic equilibria where the
  magnetic field satisfies $\nabla \times\mathbf B = \alpha \mathbf B$, where
  $\alpha$ is spatially uniform, on a periodic domain.  Using numerical
  solutions we show that generic examples of such equilibria are unstable to
  ideal modes (including incompressible ones) which are marked by exponential
  growth in the linear  phase.  We characterize the unstable mode, showing how
  it can be understood in terms of merging magnetic and current structures, and
  explicitly demonstrate its instability using the energy principle. Following
  the nonlinear evolution of these solutions, we find that they rapidly develop
  regions with relativistic velocities and electric fields of comparable
  magnitude to the magnetic field, liberating magnetic energy on dynamical
  timescales and eventually settling into a configuration with the largest
  allowable wavelength.  These properties make such solutions a promising
  setting for exploring the mechanisms behind extreme cosmic sources of gamma
  rays. 

\end{abstract}

\maketitle

{\em Introduction.}---%
Magnetic stability is a fundamental question in a range of fields from
laboratory plasma physics, where it influences the viability of fusion
devices~\cite{Freidberg1982}; to space physics, where it controls the structure
of magnetic fields within stars and planets~\cite{Shibata2011a}.  In high-energy
astrophysics, the spontaneous release of energy associated with transitions
between magnetic equilibrium states is of particular importance 
to understanding the dramatic gamma-ray activities from pulsar wind
nebulae~\cite{Blandford2014, Blandford2015},
magnetars~\cite{Goldreich1969,Thompson1995,Yu2012,Lyutikov2006}, relativistic
jets associated with active galactic
nuclei~\cite{Blandford1977,Zhang2015,2007ApJ...664L..71A,2015arXiv150204699H},
and gamma-ray bursts.  These diverse sources exhibit powerful gamma-ray flares
on timescales short compared with their light-crossing
times~\cite{Yu2012,2007ApJ...664L..71A,2015arXiv150204699H}, and seem to require
that electrons and positrons be accelerated throughout extended regions, to
energies as high as several PeV~\cite{Abdo2011, Tavani2011}.  The most dramatic
variations are likely produced in the relativistic electromagnetic outflows away
from the central engine (neutron stars or black holes), and a mechanism is
pressingly needed to explain the rapid, volumetric conversion of magnetic energy
into high energy particles and radiation.  Here, we consider whether such a
process may be triggered by magnetic instability in the outflow.  These outflows
may initially accelerate, so that they cannot be crossed by hydromagnetic waves
in an outflow timescale. However, they will eventually be decelerated when their
momentum flux decreases to that of the external medium,
bringing disconnected regions back into causal contact where they are likely to
be unstable
\footnote{
    A cosmological analogy would be when perturbations from the epoch of
    inflation are believed to have ``re-entered'' the horizon and exhibited
    gravitational instability.
    }
    . 

To understand the conditions under which a 
plasma becomes unstable, and to follow its subsequent nonlinear evolution in an
idealized setting, we focus on a model class of force-free equilibria, which we
find evolves in a manner that is both surprising on formal grounds, and highly
suggestive of the behavior of the most dramatic cosmic sources.  Force-free
solutions, where the Lorentz force vanishes, are an excellent approximation for
highly conducting and strongly magnetized plasmas, where the plasma inertia and
pressure is sub-dominant to the magnetic field, and have been used extensively
across different fields.
A particularly important class of force-free equilibria that are conjectured to
arise naturally from magnetic relaxation are the so called Taylor states, which
satisfy the Beltrami property: $\nabla \times\mathbf{B}= \alpha \mathbf{B}$
where $\alpha$ is a constant~\cite{Taylor1974}. These solutions have played an
important role not only in laboratory plasma physics~\cite{Taylor1986}, but also
in solar
physics~\cite{1984A&A...137...63H,1989A&A...225..156D,1993ApJ...417..781V},
astrophysics~\cite{Brandenburg2005}, and beyond~\cite{Marino2013}.  In this work
we focus on space-periodic equilibria as a simple, computationally tractable
setting free of the effect of confining boundaries (as in extended outflows).
Though there is a rich literature studying such solutions~\cite{Woltjer1958,
Molodensky1974, Taylor1974,Moffatt1986, Er-Riani2014}, important facts regarding
their stability have not been appreciated. Focusing on a prototypical example,
the ``ABC" solutions~\cite{arnold1965topologie} (defined below),
in~\cite{Moffatt1986} it was claimed that such solutions are stable to
incompressible perturbations (see also~\cite{Er-Riani2014}).  
Here we show that, in fact, generic periodic Beltrami magnetic fields are
linearly unstable, including to incompressible deformations. The only exceptions
we find are special cases lacking magnetic curvature, and those in the
fundamental mode or \emph{ground state}, having the lowest magnetic energy
compatible with conservation of magnetic helicity $H_M = \int{\mathbf A \cdot
\mathbf B dV}$ (where $\mathbf A$ is the magnetic vector potential). 
The instability we find is ideal, in contrast to previous studies of
dissipative effects~\cite{Horiuchi1985}.
We find
that in the nonlinear evolution, magnetic energy is indeed liberated rapidly,
giving rise to relativistic velocities and electric fields of comparable
magnitude to the magnetic fields on dynamical timescales, and eventually
allowing the system to relax to its ground state.  These solutions are therefore
a simple, but promising setting to explore the mechanisms underlying extreme
cosmic sources of gamma rays. 

In what follows, we present simulation results showing the linear-regime
instability of a range of magnetostatic equilibria, and then illustrate the
properties of the dominant unstable mode in some example cases, independently
confirming the growth rate using the energy principle. We then compare the
results found using various degrees of magnetization, discuss the nonlinear
evolution of the instability, and conclude. We use units with $c=1$ throughout.

{\em Methodology.}---%
The equilibrium magnetic fields we study are exemplified by the three-parameter
``ABC'' field~\cite{arnold1965topologie,Dombre1986} given by
\begin{align} \label{eq:ABC}
  \mathbf{B}^E = \bigl(
  &B_3\cos \alpha z-B_2\sin \alpha y, \\
  &B_1\cos \alpha x-B_3\sin \alpha z, 
  &B_2\cos \alpha y-B_1\sin \alpha x  \nonumber
  \bigr).
\end{align}
We use some particular examples of this equilibrium solution for illustrative
purposes, but also consider the more general class of Beltrami fields
\cite{Chandrasekhar1957}
%
$
  \mathbf B = \alpha \mathbf \Psi + \nabla \times \mathbf \Psi
$
%
where the potential field $\mathbf \Psi$ is any solenoidal vector field
satisfying the vector Helmholtz equation $\nabla^2 \mathbf \Psi + \alpha^2
\mathbf \Psi = 0$, so that $\mathbf \Psi$ comprises only the Fourier harmonics
whose wave-vector $\mathbf k$ has magnitude $\alpha$. These more general
configurations are constructed by choosing random vector amplitudes for the
admissible harmonics. Our computational domain is the periodic cube of length $L
= 2\pi$ (though we restore $L$ in some places for clarity).

We simulate a perfectly conducting, magnetized fluid 
and consider cases with different finite values of the volume-averaged
magnetization parameter $\sigma:=\langle B^2/4\pi\rho h \rangle$ where $\rho h$
is the fluid enthalpy (treated using the ideal relativistic
magnetohydrodynamic equations and the code in \cite{Zrake2011}), as well as
the limiting case of a completely magnetically dominated plasma,
$\sigma=\infty$ (treated by force free electrodynamics~\cite{1997PhRvE..56.2181U,
Thompson1998,Pfeiffer2013}).  See \emph {supplemental
material} below~\nocite{Blandford2002luml.conf..381B,2010PhRvD..81h4007P,kreiss1973,Gruzinov1999astro.ph..2288G}
\newcommand{\supplementalmaterial}{See Supplemental Material url, which includes Refs.~\cite{Blandford2002luml.conf..381B,2010PhRvD..81h4007P,kreiss1973,Gruzinov1999astro.ph..2288G}}
for details.

\begin{figure}
  \begin{center}
    \includegraphics[width=\columnwidth,draft=false]{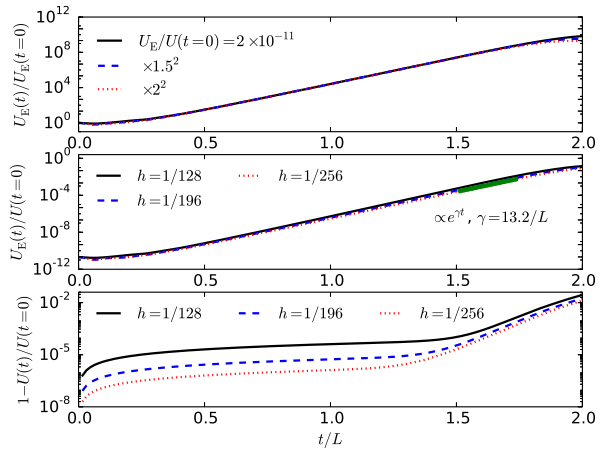}
  \end{center}
  \caption{
    Results from simulations with $\ta^2=11$. Top: The growth of the electric
    field energy $U_{\rm E}$, normalized by its initial value, for three
    different values of the initial perturbation.  Middle: The growth in $U_{\rm
    E}$ for three different resolutions, along with an exponential fit. The
    difference between the best fit exponent for the high resolution, and the
    Richardson extrapolated value using all three resolutions, is
    $\approx0.1\%$, the extrapolation being consistent with between first and
    second-order convergence.  The bottom panel illustrates the conservation of
    total energy $U$ for three different resolutions.  Though initially
    higher-order when the equilibrium-solution truncation error dominates,
    the convergence eventually drops to first-order, presumably because (as
    discussed below) the unstable solution has non-smooth features.
    Conservation of magnetic helicity is similar.  \label{ksq11_comp}
  }
\end{figure}

{\em Instability in the Linear Regime.}---%
For this class of magnetic equilibria, we find generic solutions with $\ta^2>1$
to be unstable to linear ideal perturbations that are characterized by
exponential growth of the electric field.  Fig.~\ref{ksq11_comp} illustrates
this for a case with $\ta^2=11$. (Here we present results from $\sigma=\infty$ 
simulations, and in a later section we compare these to finite magnetization
cases.) The magnitude of
the growing solution is proportional to the initial
perturbation\footnote{Here the initial perturbation we use is an
electric field with $E^{x}=E_0\cos(2\pi y/L)$ and where the other
components are given by cyclic permutations of $\{x,y,z\}$. From this we
subtract out any component parallel to $\mathbf{B}$.}, consistent with a
linear instability. The growth rate of, e.g., the electric field energy is
converging to $\gamma\approx4.0\ta/L$ with increased resolution, evidence that
the instability is not due to numerical/non-ideal effects.

\begin{figure}
  \begin{center}
    \includegraphics[width=\columnwidth,draft=false]{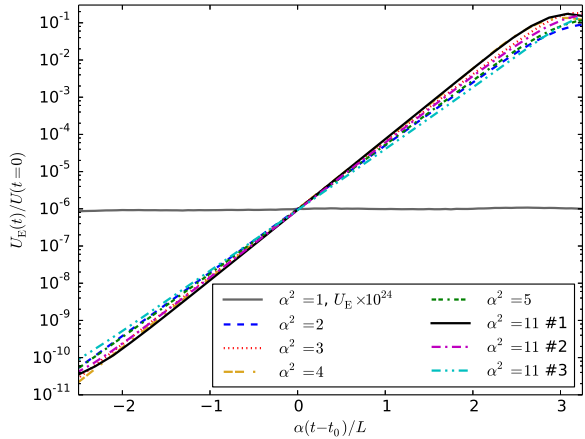}
  \end{center}
  \caption{ The growth in electric field energy for various values of $\ta^2$
  and three different realizations for $\ta^2=11$. Time has been shifted so that
  all the curves have an abscissa of 0 at the ordinate value of $10^{-6}$, and
  the time axis has been scaled by $\ta$ which gives the different examples
  roughly the same slope. The $\ta^2=1$ simulation does not exhibit exponential
  growth, and has been scaled up by an overall factor.
    \label{ksq_comp}
  }
\end{figure}

Other equilibrium solutions exhibit similar exponentially growing solutions, as
shown for some example cases in Fig.~\ref{ksq_comp}.  This holds for wavelengths
larger than the fundamental mode for the domain, $\ta^2=1$, which is known to be
stable~\cite{Moffatt1986, Er-Riani2014}.  The growth rate of the instability is
also roughly proportional to $\ta$, though there is dependence on the particular
realization used.

\begin{figure*}
  \begin{center}
    \includegraphics[width=0.66\columnwidth,draft=false]{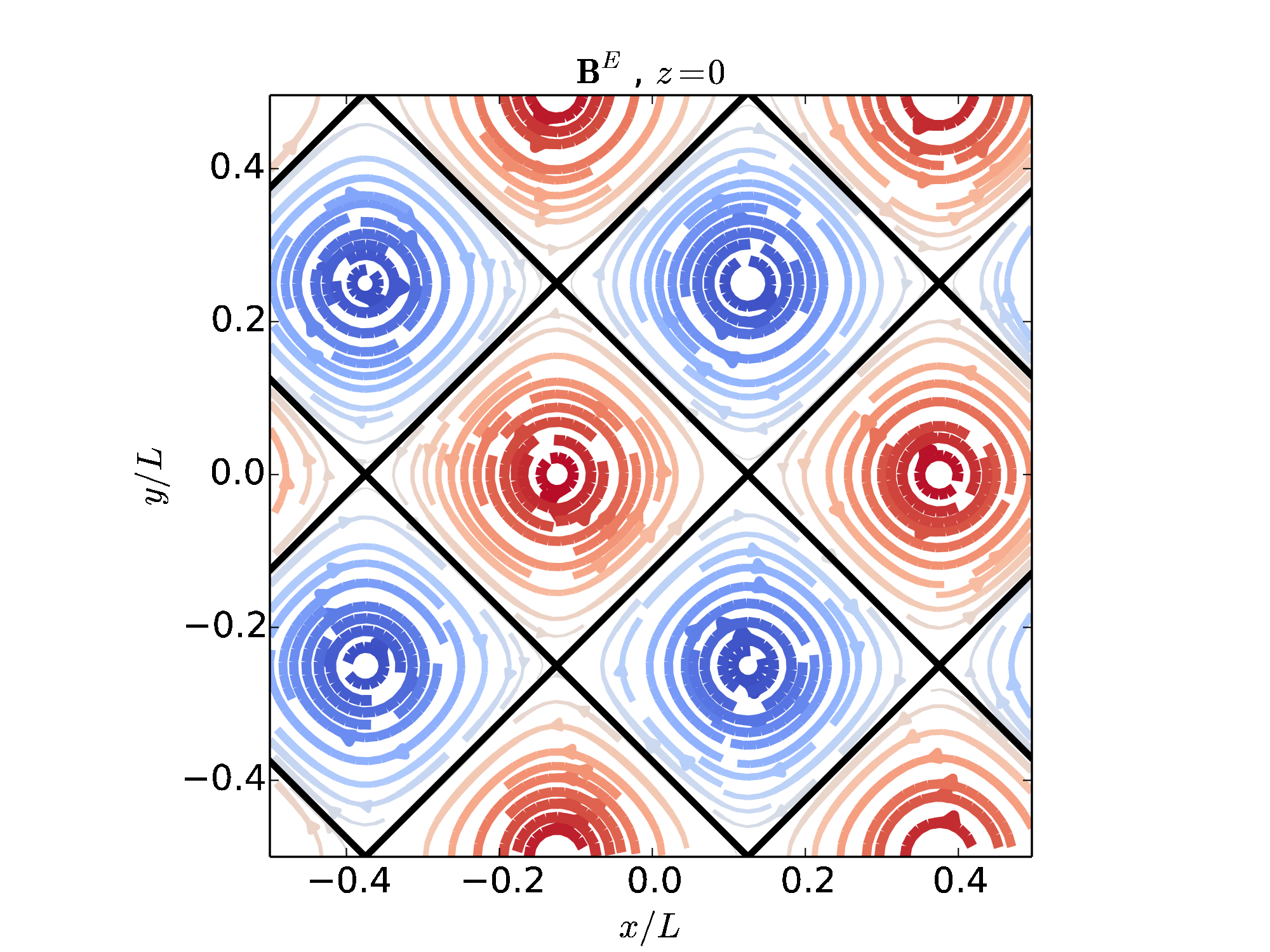}
    \includegraphics[width=0.66\columnwidth,draft=false]{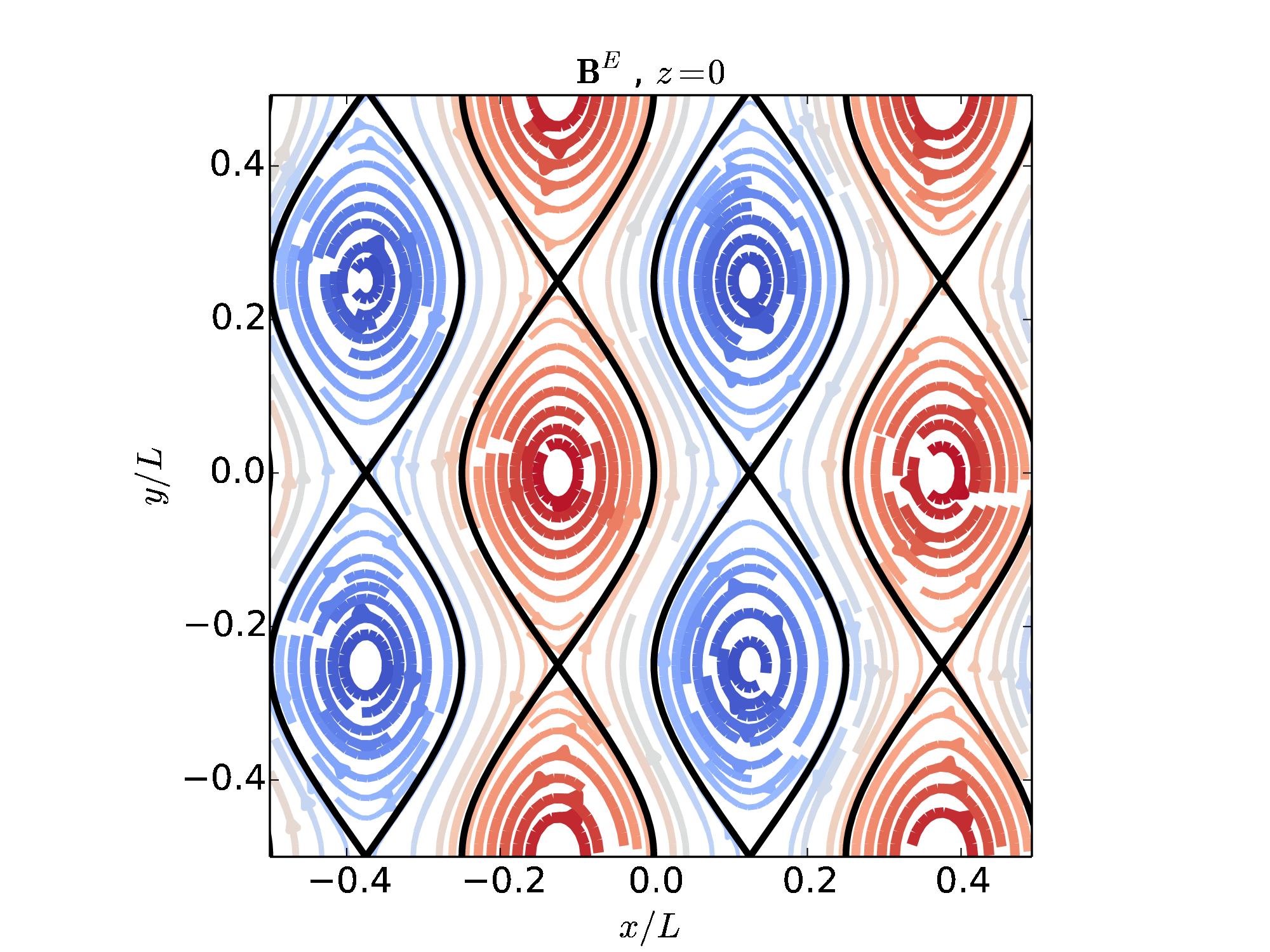}
    \includegraphics[width=0.66\columnwidth,draft=false]{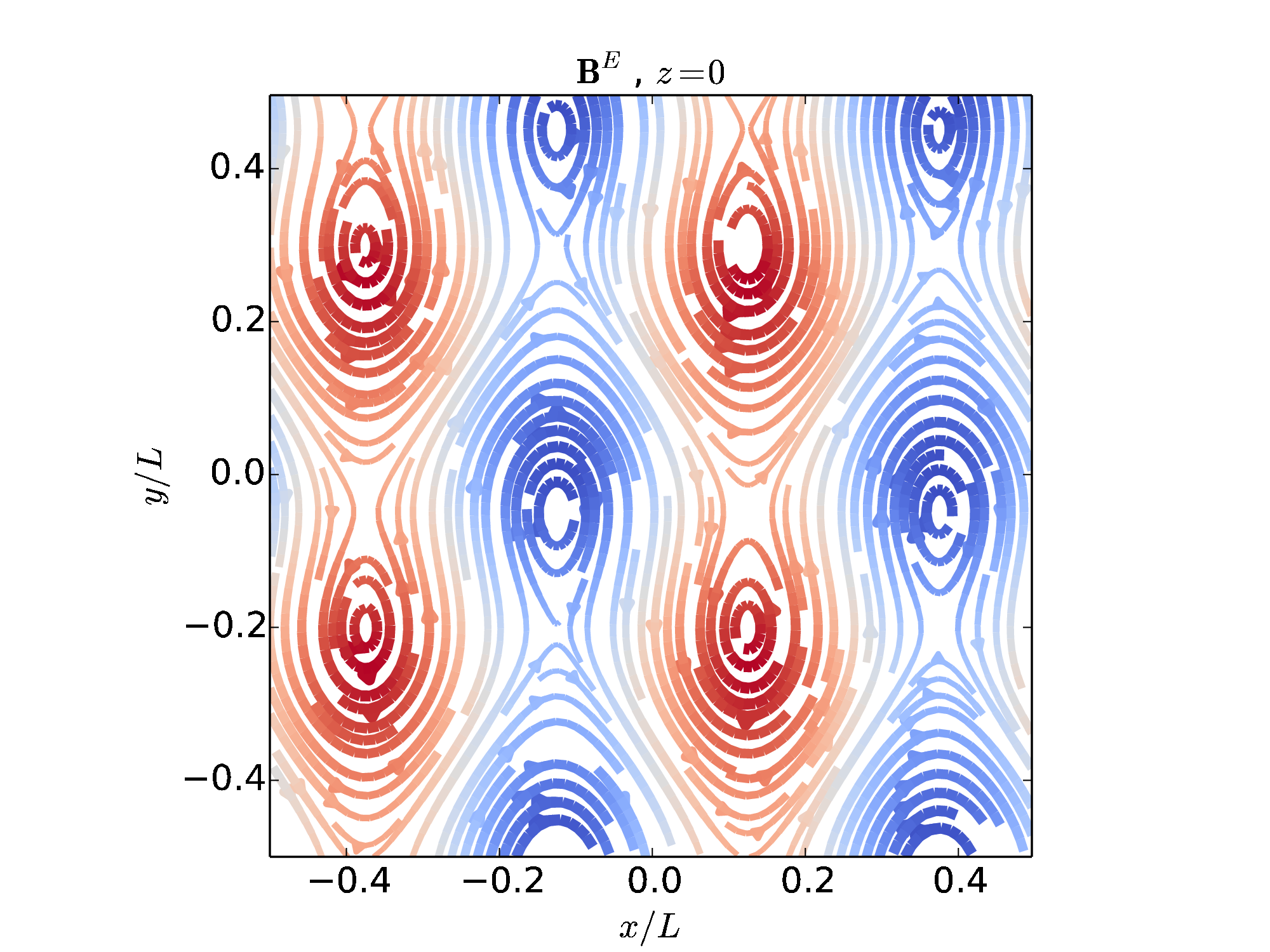}
    \includegraphics[width=0.66\columnwidth,draft=false]{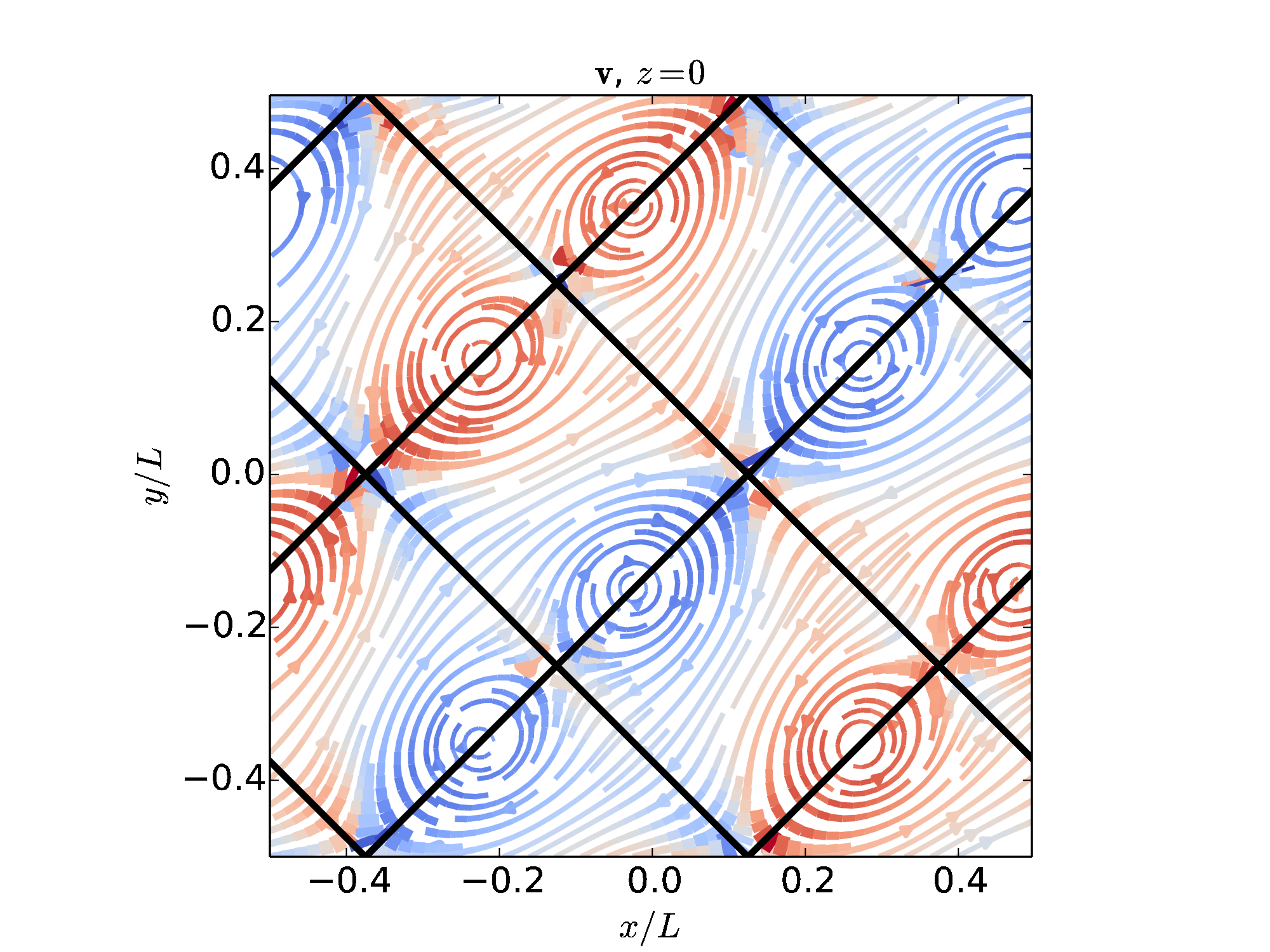}
    \includegraphics[width=0.66\columnwidth,draft=false]{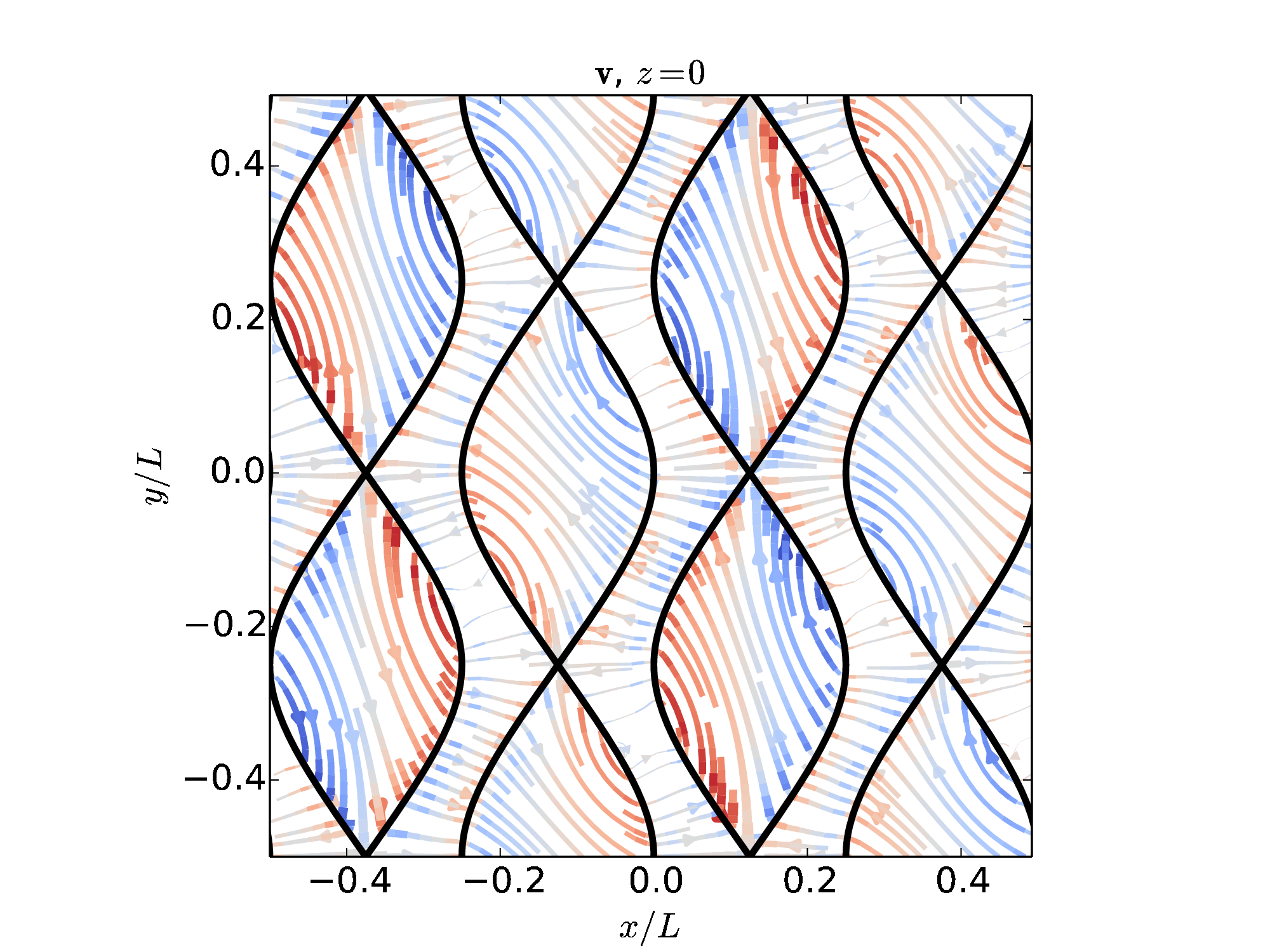}
    \includegraphics[width=0.66\columnwidth,draft=false]{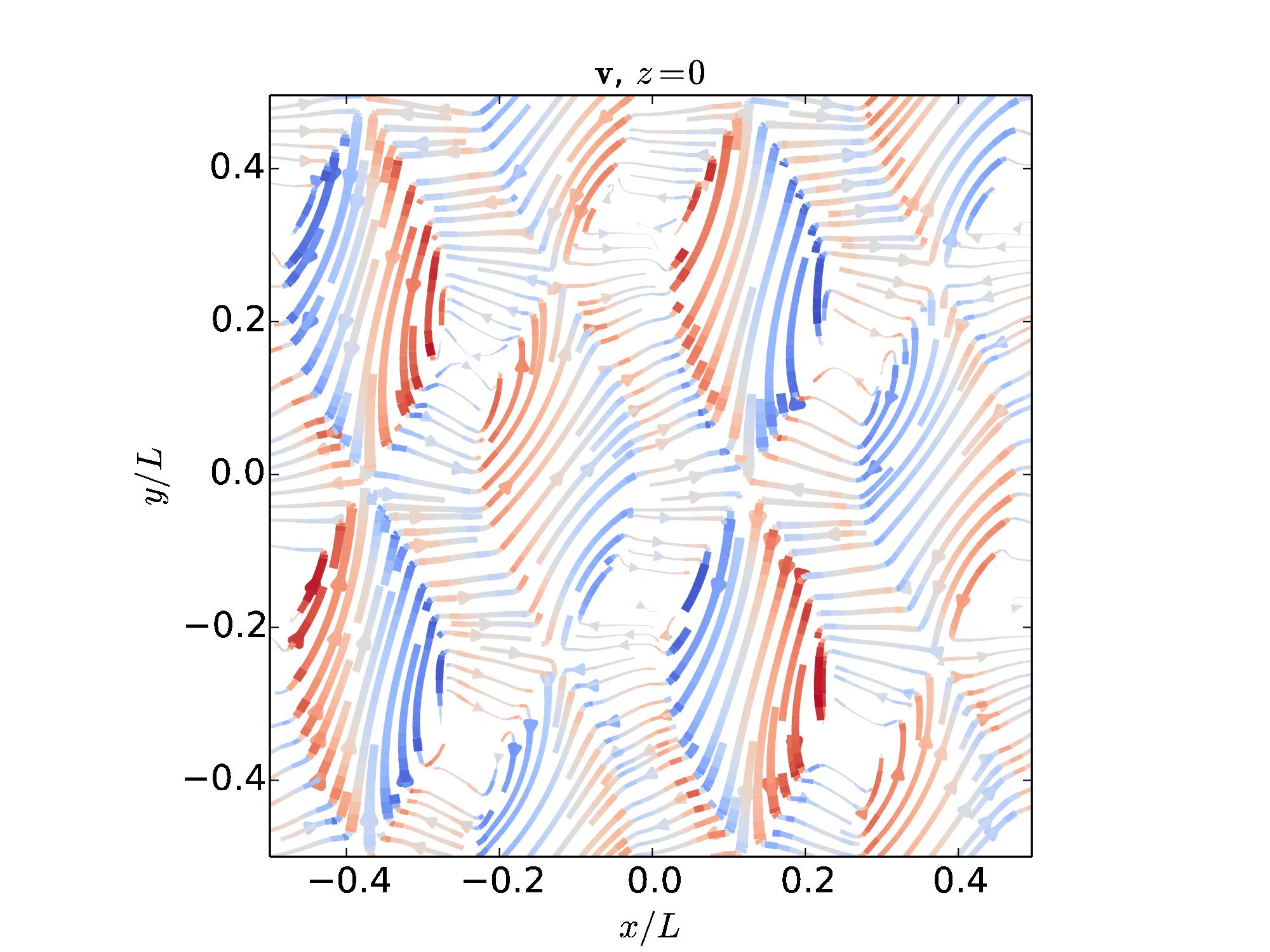}
  \end{center}
  \caption{ Streamlines of a magnetic field equilibrium solution $\mathbf{B}^E$
    given by Eq.~\ref{eq:ABC} with $\ta=2$ and various coefficients (top), and
    the corresponding velocity field $\mathbf{v}=\mathbf{E}\times
    \mathbf{B}^E/|\mathbf{B}^E|^2$ of the unstable mode arising from the
    simulations (bottom) in the $z=0$ plane.  The equilibrium solutions, from
    left to right, correspond to $(B_1,B_2,B_3)=(1,1,0)$, $(1,1/2,0)$, and 
    $\approx(-0.814, 0.533, 0.232)$, respectively. The
    color indicates the perpendicular vector component with red and blue
    representing, respectively, out of the page and into the page.  The
    thickness of the streamline is proportional to the vector magnitude. The
    black lines indicate the location of the separatrices in the equilibrium
    solutions.  \label{streamplots} }
\end{figure*}

{\em The Dominant Unstable Mode.}---%
In order to illustrate the nature of the instability, we focus on a simple type
of $\ta=2$ equilibrium solution given by Eq.~\ref{eq:ABC}.  We illustrate this
solution for three different choices of coefficients in Fig.~\ref{streamplots}.
As discussed in~\cite{Dombre1986} for the mathematically equivalent Euler flow,
these solutions have a rich structure. The $(B_1,B_2,B_3)=(1,1,0)$ case consists
of ``vortices:'' regions of helical field (and current) lines circling a central
axis. The $(B_1,B_2,B_3)=(1,1/2,0)$ case has vortices as well as ``shear
layers:" wavy field lines that begin and end on opposite sides of the domain. In
addition to these two cases with $z$-translational symmetry, we also show a more
generic case where all three coefficients are nonzero (and that like the second
case, has no places where $\mathbf{B}^E=0$).

In Fig.~\ref{streamplots} we also show the corresponding velocity field
$\mathbf{v}=\mathbf{E}\times \mathbf{B}^E/|\mathbf{B}^E|^2$ (which will be
proportional to the displacement $\mathbf{\xi}$ for an eigenmode) characterizing
the dominant instability arising in each case.  This is calculated from a
numerical snapshot after the instability has grown by roughly 10 orders of
magnitude --- seeded in this case just by truncation error --- but is still in
the linear regime ($|\mathbf{E}|\sim 10^{-4} |\mathbf{B}^E|$).  The velocity
field acts to bring together vortices, or current channels, circulating in the
same direction in order to move towards a larger wavelength, lower magnetic
energy configuration.  The nonlinear evolution is characterized by the merging
of magnetic vortices and can thus be related to the coalescence instability of
magnetic islands~\cite{Finn1977}.  We can also see that the velocity field
appears to have non-smooth features, reminiscent of spontaneous current
sheets~\cite{Parker1994}, that occur at the separatrices dividing the vortices
and shear layers.  Though the generic case lacks $z$-translational symmetry, it
appears qualitatively similar to the second case.

From the $(B_1,B_2,B_3)=(1,1,0)$ to the $(1,1/2,0)$ case, the growth rate of the
instability decreases by a factor of $\approx1.9$ with the addition of the shear
layers in the equilibrium solution.  In fact, the growth rate decreases
monotonically with $B_2$, and as $B_2\to0$ and the vortices shrink to zero
volume, the growth rate of the instability also goes to zero. In fact, it can be
shown (see supplemental material below) that the single mode solutions are all
stable, including those at short wavelengths.

For the generic case where the three coefficients are nonzero, we directly
confirm that the instability is linear and ideal by using the energy
principle~\cite{Bernstein1958}. This implies that an equilibrium solution
$\mathbf{B}^E$ satisfying the Beltrami property is unstable to a displacement
$\mathbf{\xi}$ if
\beq
     \omega^2 := \frac{\int \bigl[|\delta \mathbf{B}|^2
     -\alpha (\mathbf{\xi}\times \mathbf{B}^E)\cdot \delta \mathbf{B}) \bigr]dV}
    {\int |\mathbf{B}^E|^2|\mathbf{\xi}|^2dV}<0 ,
 \label{eq:omega}   
\eeq
where $\delta \mathbf{B}:=\nabla\times(\mathbf{\xi}\times \mathbf{B}^E)$, and
that the instability should grow at least as fast as $|\omega|$. Computing this
quantity using finite differences, with the numerical velocity field, gives a
value of $|\omega| L\approx 2.4$, near the growth rate of $2.6$ measured for the
electric field in the simulation.

Although the unstable displacement from the simulations has both non-smooth and
compressible ($\int |\nabla \cdot \mathbf{v}|dV\approx 0.15 \int |\nabla \times
\mathbf{v}|dV$) features, these are not necessary for instability. For example,
by applying a low-pass filter to the Fourier harmonics of $\mathbf{\xi}$ one can
construct an ideal perturbation that has no power at scales $|\tk|>29$,
and yet gives a value of $|\omega|L\approx2.3$ upon evaluating
Eq.~(\ref{eq:omega}) algebraically in $k$-space.  Furthermore, we have also
explicitly constructed (see supplemental material below) analytical examples of
smooth and incompressible displacements that destabilize particular 
Beltrami solutions,
which are counterexamples to the claims in \cite{Moffatt1986}.

\begin{figure}
  \begin{center}
    \includegraphics[width=\columnwidth,draft=false]{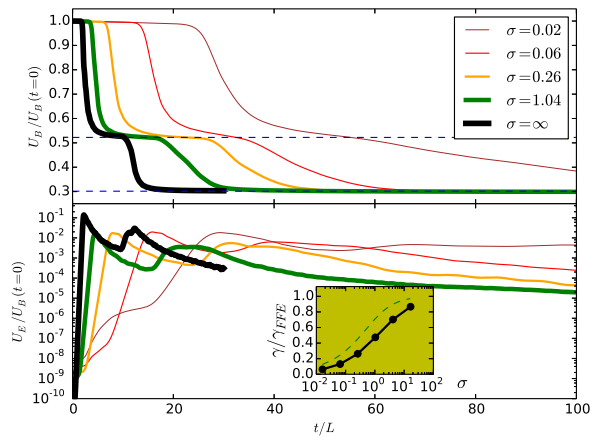}
  \end{center}
  \caption{A comparison of the decay of an $\ta^2=11$ equilibrium in 
  simulations with different values of magnetization parameter $\sigma$.
  Shown is the magnetic energy (top) and kinetic/electric field energy 
  (bottom). The  horizontal dashed lines in the top panel indicate the magnetic
  energy of $\ta^2=3$ and $\ta^2=1$ states with the same helicity. The bottom
  inset shows the linear growth rate $\gamma$ measured for runs having different
  magnetization parameters, along with the Alfv\'{e}n speed (dashed line) for
  comparison. 
    \label{mhd_ff_comp}
  }
\end{figure}

{\em Magnetization comparison and nonlinear evolution.}---%
The same qualitative behavior is observed when the plasma is evolved with
different magnetization parameters.  Fig.~\ref{mhd_ff_comp} shows the evolution
of the kinetic and magnetic energy $U_{B}$ for different parameters $\sigma$
values, including the limiting case of $\sigma=\infty$. The lower inset of
Fig.~\ref{mhd_ff_comp} shows that the growth rate of the unstable mode increases
monotonically with increasing $\sigma$ and is roughly proportional to the
Alfv\'{e}n speed $v_A=\sqrt{\sigma/(1+\sigma)}$.

In both finite and infinite magnetization cases, exponential growth of the
unstable displacement persists until its velocity $|\mathbf v|$ approaches the
Alfv\'{e}n speed $v_A$ (near the speed of light for large $\sigma$) and
higher-order evolutionary terms dominate. Following the turbulent state, the
system settles into a lower energy equilibrium. Somewhat surprisingly, we
\emph{do not} observe direct evolution into the lowest energy ($\ta = 1$) state
for all cases. For example, the $\ta^2 = 11$ state in Fig.~\ref{mhd_ff_comp}
first transitions into a configuration with $\approx97\%$ of its spectral energy
in modes with $\tk^2 = 3$, where it remains for about ten Alfv\'{e}n crossing
times, before making a second transition into the ground state, where 99\% of
its energy in $\tk^2 = 1$ modes. The lifetime of the intermediate state may be
related to the fact that $\approx88\%$ of the energy is in a single $\tk^2=3$
mode.

During the nonlinear evolution, regions develop where $|\mathbf{E}|$ is
comparable, and even exceeds $|\mathbf{B}|$.  Since the hyperbolicity of the
equations breaks down for $\sigma=\infty$ when this happens, to evolve further
we handle such regions with an ad-hoc prescription where we simply reduce the
electric-field magnitude to equal that of the magnetic field, leading to a
reduction of energy.  The finite $\sigma$ cases do not suffer from this problem,
and our scheme explicitly conserves total energy, but permits conversion of
magnetic or kinetic energy into internal energy, especially at shocks or places
where the magnetic field is nearly discontinuous.  Encouragingly, we still find
consistency between these different (and somewhat arbitrary) types of energy
dissipation in the non-linear regime.  For example, as shown in
Fig.~\ref{mhd_ff_comp}, we find the same energy levels associated with the
intermediate and final magnetic equilibria.  This is consistent with
conservation of magnetic helicity. Since the Beltrami fields have $\mathbf
B=\alpha\mathbf A$, their helicity is $2U_B/\alpha$, and the ratio of magnetic
energy in the $\alpha_i$ and $\alpha_f$ equilibria is simply $\alpha_f /
\alpha_i$.  Accordingly, we do not expect the dissipation mechanism to have much
influence on the energy of the final state if helicity is preserved.  (For the
simulations shown in Fig.~\ref{mhd_ff_comp}, $H_M$ is constant to $\sim0.1\%$.)
But understanding the details of the energy dissipation 
will require better physical modeling.

{\em Conclusions.}---%
We studied periodic Beltrami magnetic fields and found that they were unstable
to ideal modes.  Though we focused on the relativistic case, since the
instability is linear in velocity, it also applies to the nonrelativistic
setting. This contrasts with~\cite{Moffatt1986} --- where it was
concluded that such solutions are linearly stable against incompressible
perturbations --- and suggests that the relaxation of a complex magnetic field will
\emph{not} terminate at a small wavelength equilibrium, but instead undergo a
so-called inverse helicity cascade where magnetic energy goes to the largest
available scale~\cite{Frisch1975, Alexakis2006, Zrake2014, Brandenburg2015}.
There are known examples of unstable cylindrically-symmetric Beltrami
solutions~\cite{Voslamber1962}, and studying a broader class of 
geometries would be interesting follow-up work.

For highly magnetized, relativistic plasma, the instability gives rise to
regions where the electric field magnitude is comparable to the magnetic field
on dynamical timescales. In extreme cosmic sources of gamma rays, where such
configurations may be relevant, these would be likely sites of particle
acceleration and photon emission.  Understanding the details of this, including
the role of magnetic
reconnection~\cite{2005A&A...441..845D,2008ApJ...677..530Z,2014ApJ...780....3U}
and turbulence~\cite{1999ApJ...517..700L} in ultimately dissipating energy, and
determining the nature of the acceleration
mechanism~\cite{2010MNRAS.408L..46G,2011ApJ...735..102K,2014ApJ...783L..21S},
will require kinetic simulations incorporating radiative losses, something we
plan for future work.  

We thank Antony Jameson, as well as the anonymous referees for useful
suggestions.  YY and RB thank Keith Moffatt for helpful advice and
encouragement.  This work was supported in part by the U.S. Department of Energy
contract to SLAC no.  DE-AC02-76SF00515, NSF grant AST 12-12195, as well as the
Simons    Foundation, the Humboldt Foundation, and the Miller Foundation (RB).
YY gratefully acknowledges support from the KIPAC Gregory and Mary Chabolla
fellowship and the Gabilan Fellowship awarded by Stanford University.
Simulations were run on the Bullet Cluster at SLAC, the Sherlock Cluster at
Stanford University, and using resources provided by the NASA High-End Computing
(HEC) Program through the NASA Advanced Supercomputing (NAS) Division at Ames
Research Center.

\onecolumngrid
\appendix

\section{Supplemental material: Details of numerical methods}
In the main text we assume a perfectly conducting medium and perform simulations of cases having various
degrees of magnetic dominance.  The limiting case of a completely magnetically
dominated plasma is treated by FFE~\cite{1997PhRvE..56.2181U,
Thompson1998,Pfeiffer2013}.  The evolution equations of FFE are just the Maxwell
equations with a prescription for the current derived from the assumption that
the Lorentz force vanishes and can be written as \cite{Blandford2002luml.conf..381B, Pfeiffer2013} (we use Heaviside-Lorentz units and set $c=1$ throughout):
\begin{subequations}\label{eq:FF_evo}
\begin{gather}
\frac{\partial \mathbf{B}}{\partial t}=-\nabla \times \mathbf{E},\\
\frac{\partial \mathbf{E}}{\partial t}=\nabla \times \mathbf{B}-\mathbf{j},\\
\mathbf{j}=\frac{1}{B^2}\left[\left(\mathbf{B}\cdot (\nabla \times \mathbf{B})-\mathbf{E}\cdot (\nabla \times \mathbf{E})\right)\mathbf{B}+(\nabla \cdot \mathbf{E})\mathbf{E}\times \mathbf{B}\right].
\end{gather}
\end{subequations}
We numerically solve these using fourth-order
finite difference stencils and Runge-Kutta time stepping. 
We use hyperbolic divergence cleaning to exponentially damp violations of the
$\nabla \cdot \mathbf{B}=0$ constraint as in~\cite{2010PhRvD..81h4007P}. 
The $\mathbf{E}\cdot\mathbf{B}=0$
constraint is explicitly enforced by redefining $\mathbf{E} \rightarrow
\mathbf{E}-\mathbf{B} (\mathbf{E} \cdot \mathbf{B})/B^2$ at every coarse time
step.  We apply standard sixth-order Kreiss-Oliger~\cite{kreiss1973} numerical dissipation to all the hyperbolic variables to suppress high frequency numerical error.  The FFE code is
parallelized using the PAMR/AMRD software library~\footnote{http://laplace.physics.ubc.ca/Group/Software.html}.

We also solve the ideal relativistic magnetohydrodynamic (RMHD) equations (with a
$\Gamma=4/3$ equation of state) for different values of the volume-averaged
magnetization parameter $\sigma:=\langle B^2/4\pi\rho h \rangle$ (where $\rho h$
is the fluid enthalpy).  In addition to the specified magnetic field,
we use a restmass density and pressure that are initially equal and uniform.
We use a second-order, constrained-transport, finite-volume
scheme that explicitly conserves mass, energy, momentum, and magnetic flux.  
Full details of the code are described in \cite{Zrake2011}.

\section{Supplemental material: Unstable modes with analytical methods}
In this sections, we verify the existence of unstable ideal modes
for periodic Beltrami solutions using analytical methods.
Introducing $\mathbf{v}=\mathbf{E}\times\mathbf{B}/B^2$ and
$\mathbf{E}=-\mathbf{v}\times\mathbf{B}$, the evolution
equations~(\ref{eq:FF_evo}) can be rewritten in
terms of the following equations \cite{Gruzinov1999astro.ph..2288G}
\begin{subequations}\label{eq:FF_evo1}
\begin{gather}
\frac{\partial \mathbf{B}}{\partial t}=\nabla \times (\mathbf{v}\times \mathbf{B}),\\
\frac{\partial }{\partial t}\left(B^2\mathbf{v}\right)=(\nabla \times \mathbf{B})\times \mathbf{B}+(\nabla \times \mathbf{E})\times \mathbf{E}+(\nabla \cdot \mathbf{E})\mathbf{E}
\end{gather}
\end{subequations}

Now consider small perturbation on a static equilibrium state which satisfies $\nabla \times \mathbf{B}_0=\alpha \mathbf{B}_0$ with $\alpha$ being a constant. Let $\mathbf{B}=\mathbf{B}_0+\mathbf{B}_1$, and define the displacement field $\vec{\xi}$ such that to first order $\partial\vec{\xi}/\partial t=\mathbf{v}$. Suppose the perturbation can be decomposed into normal modes $\propto e^{i\omega t}$. The linearized version of Equation (\ref{eq:FF_evo1}) now becomes
\begin{subequations}\label{eq:FF_linear}
\begin{gather}
\mathbf{B}_1=\nabla \times (\vec{\xi}\times \mathbf{B}_0),\\
\begin{split}
-\omega^2B^2\vec{\xi}&=(\nabla \times \mathbf{B}_1-\alpha\mathbf{B}_1)\times \mathbf{B}_0\\
&=(\nabla \times \mathbf{B}_0)\times (\nabla \times (\vec{\xi}\times \mathbf{B}_0))+[\nabla \times (\nabla \times (\vec{\xi}\times \mathbf{B}_0))]\times \mathbf{B}_0\\
&\equiv\hat{\mathbf{K}}\cdot \vec{\xi} 
\end{split}
\end{gather}
\end{subequations}
In the equation for $\vec{\xi}$ the differential operator $\hat{\mathbf{K}}$ is self-adjoint under periodic boundary conditions \cite{Bernstein1958}, so we can define a potential energy 
\begin{equation}\label{eq:potential energy}
V=-\frac{1}{2}\int\! d^3x\, \vec{\xi} \cdot \hat{\mathbf{K}}\cdot \vec{\xi} = \frac{1}{2} \int\! d^3x\left\{[\nabla \times (\vec{\xi} \times \mathbf{B})]^2-\alpha (\vec{\xi} \times \mathbf{B})\cdot [\nabla\times (\vec{\xi} \times \mathbf{B})]\right\}
\end{equation}
As a result
\begin{equation}\label{eq:variational}
    \omega ^2=\frac{-\int\! d^3x\, \vec{\xi} \cdot \hat{\mathbf{K}}\cdot \vec{\xi} }{\int\! d^3x\, B^2 \vec{\xi}_{\perp}\cdot \vec{\xi}_{\perp}}=\frac{V}{\frac{1}{2}\int\! d^3x\, B^2 (\vec{\xi}_{\perp})^2}.
\end{equation}
This allows a variational approach to the stability problem: we can use trial functions to get an upper limit on the lowest $\omega^2$. An equilibrium state is unstable as long as we can find one trial function that renders $V$ negative. In the following we show a few examples.

\subsection{1D equilibria}
One-dimensional force-free equilibria in periodic box can be written as $\mathbf{B}_0=(\cos\psi(z),\sin\psi(z),0)$, where $\psi(z)=\alpha z$ for linear force-free fields. We can write the normal mode perturbation as $\vec{\xi }=\vec{\xi }(z)e^{i \left( k_xx+k_yy-\omega  t\right)}$. Since the component of $\vec{\xi}$ parallel to the background $\mathbf{B}_0$ does not have physical significance, we can impose the requirement $\vec{\xi}\perp\mathbf{B}_0$. Then from Equation (\ref{eq:FF_linear}), we get
\begin{subequations}\label{eq:Maxwell_covariant}
\begin{gather}
\xi _x(z)=-\frac{i \sin  \psi  \left(k_y \cos  \psi -k_x\sin  \psi \right)}{k_x^2+k_y^2-\omega ^2}\xi _z'(z),\\
\frac{1}{\omega ^2-\left(k_x\cos  \psi +k_y\sin  \psi \right){}^2}\frac{d}{\text{dz}}\left(\left(\omega ^2-\left(k_x\cos  \psi +k_y\sin  \psi \right){}^2\right)\frac{\text{d$\xi $}_z}{\text{dz}}\right)+\left(\omega ^2-k_x^2-k_y^2\right)\xi _z=0
\end{gather}
\end{subequations}
Eigensolutions with $\omega ^2>k_x^2+k_y^2$ exist and they correspond to oscillating normal modes. However, we do not have normal modes with $\omega^2<0$: otherwise $\omega ^2-\left(k_x\cos  \psi +k_y\sin \psi\right){}^2<0$ and the solution would be essentially exponential so it cannot satisfy the boundary condition. As a result, the 1D equilibria with periodic boundary conditions are stable to ideal modes.

\subsection{2D and 3D equilibria}
In this case it's not realistic to solve the normal mode equation so we make use of the variational principle. In periodic box it's convenient to use Fourier basis to construct our parameterized trial functions: $\vec{\xi}=\sum_{m}\vec{\xi}_{m}e^{i\mathbf{k}_m\cdot\mathbf{x}}$, where $\mathbf{k}_{-m}=-\mathbf{k}_m$, $\vec{\xi}_{-m}=\vec{\xi}_{}^*$ to ensure reality. $\mathbf{B}_0$ can also be decomposed into Fourier components: $\mathbf{B}_0=\sum_{n}\mathbf{B}_{n}e^{i\mathbf{a}_n\cdot \mathbf{x}}$, where $\mathbf{a}_{-n}=-\mathbf{a}_n$, $\mathbf{B}_{-n}=\mathbf{B}_{n}^*$, $\mathbf{a}_n\cdot\mathbf{B}_{n}=0$, $|\mathbf{a}_n|=|\alpha|$ and $i\mathbf{a}_n\times\mathbf{B}_{n}=\alpha\mathbf{B}_{n}$. Then the integrals for calculating the potential energy in Equation (\ref{eq:variational}) only involve algebraic manipulations (cf \cite{Moffatt1986}, with modifications):
\begin{equation}
\begin{split}
V&=\frac{1}{2}\int_0^{2\pi}\int_0^{2\pi}\int_0^{2\pi}\mathbf{B}_1\cdot[\mathbf{B}_1-\vec{\xi}\times(\nabla\times\mathbf{B}_0)]\,dx dy dz \equiv\langle\mathbf{B}_1\cdot[\mathbf{B}_1-\vec{\xi}\times(\nabla\times\mathbf{B}_0)]\rangle\\
&=\sum_{\substack{\mathbf{K}=\mathbf{k}_m+\mathbf{a}_n\\ =\mathbf{k}_m'+\mathbf{a}_n'}}\sum_n\sum_{n'}[(\mathbf{k}_{m'}+\mathbf{a}_{n'})\times(\vec{\xi}_{m'}^*\times\mathbf{B}_{n'}^*)] \cdot [(\mathbf{k}_{m}+\mathbf{a}_{n})\times(\vec{\xi}_{m}\times\mathbf{B}_{n})-\vec{\xi}_m\times(\mathbf{a}_n\times\mathbf{B}_n)]\\
&=\sum_{\mathbf{K}=\mathbf{k}_m+\mathbf{a}_n}\left\{\frac{1}{2}\left|\sum_n[(\mathbf{k}_m\cdot\mathbf{B}_n)\vec{\xi}_m-(\mathbf{k}_m\cdot\vec{\xi}_m)\mathbf{B}_n]\right|^2 -\frac{1}{2}\left|\sum_n(\vec{\xi}_m\cdot\mathbf{B}_n)\mathbf{a}_n\right|^2 -\frac{1}{2}\left|\sum_n(\vec{\xi}_m\cdot\mathbf{a}_n)\mathbf{B}_n\right|^2 \right.\\ &+\left.\frac{1}{2}\left|\sum_n[(\mathbf{k}_m\cdot\mathbf{B}_n)\vec{\xi}_m-(\mathbf{k}_m\cdot\vec{\xi}_m)\mathbf{B}_n-(\vec{\xi}_m\cdot\mathbf{B}_n)\mathbf{a}_n -(\vec{\xi}_m\cdot\mathbf{a}_n)\mathbf{B}_n]\right|^2\right\}
\end{split}
\end{equation}
For certain equilibria this expression is not positive definite and can turn out to be negative using appropriate trial functions. 

In the following we focus on a particular case, the ABC field
\begin{equation}\label{eq:ABCnew}
    \mathbf{B}=B_1(0, \cos  \alpha  x, -\sin  \alpha  x)+B_2(-\sin  \alpha  y,0,\cos  \alpha  y)+B_3(\cos  \alpha  z, -\sin  \alpha  z,0),
\end{equation}
which has been studied in the main text using numerical simulations. 

\begin{figure}[htb]
  \centering
  \subfigure[]
    {
        \includegraphics[width=0.3\textwidth]{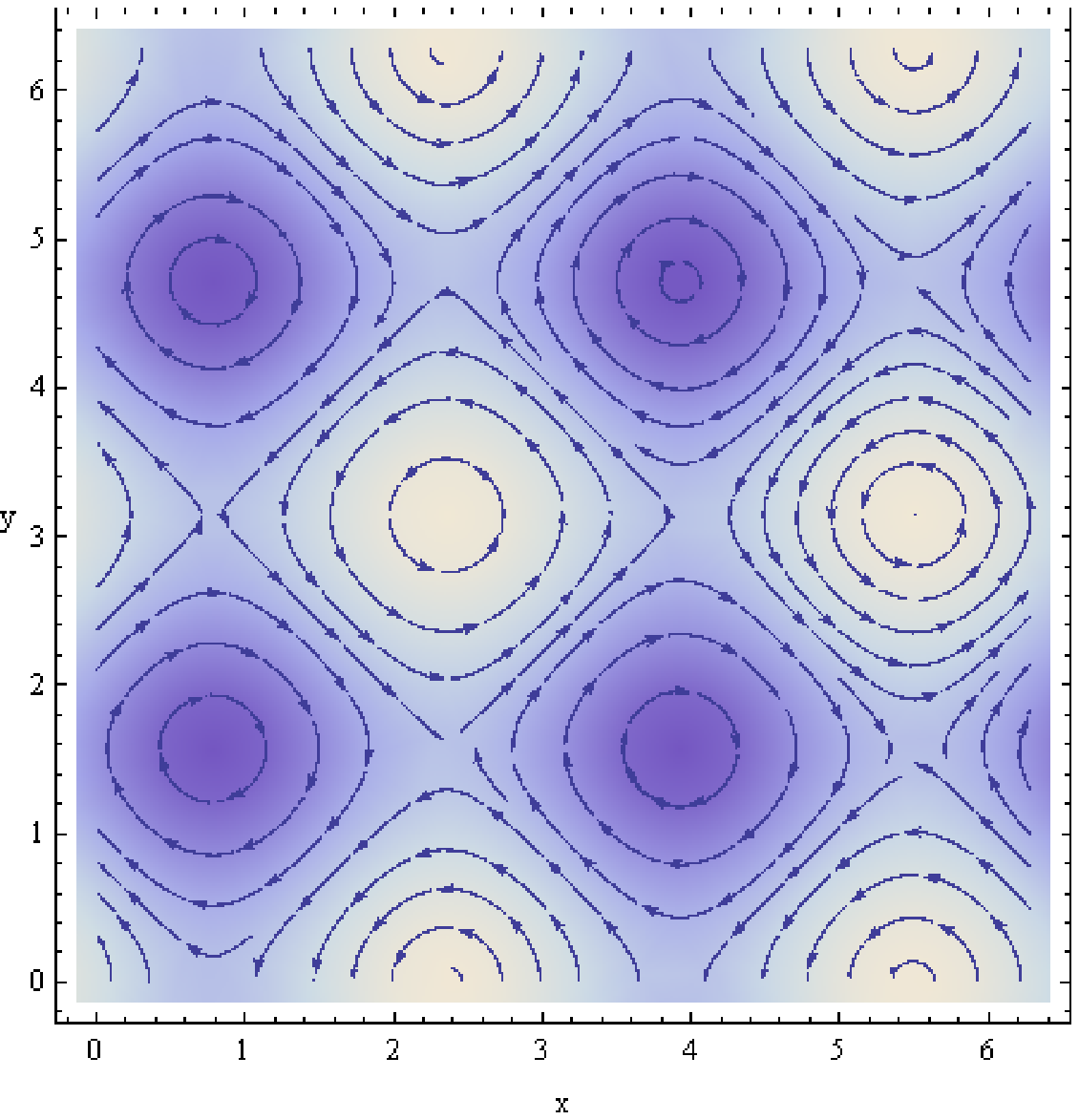}
    }
    \subfigure[]
    {
        \includegraphics[width=0.3\textwidth]{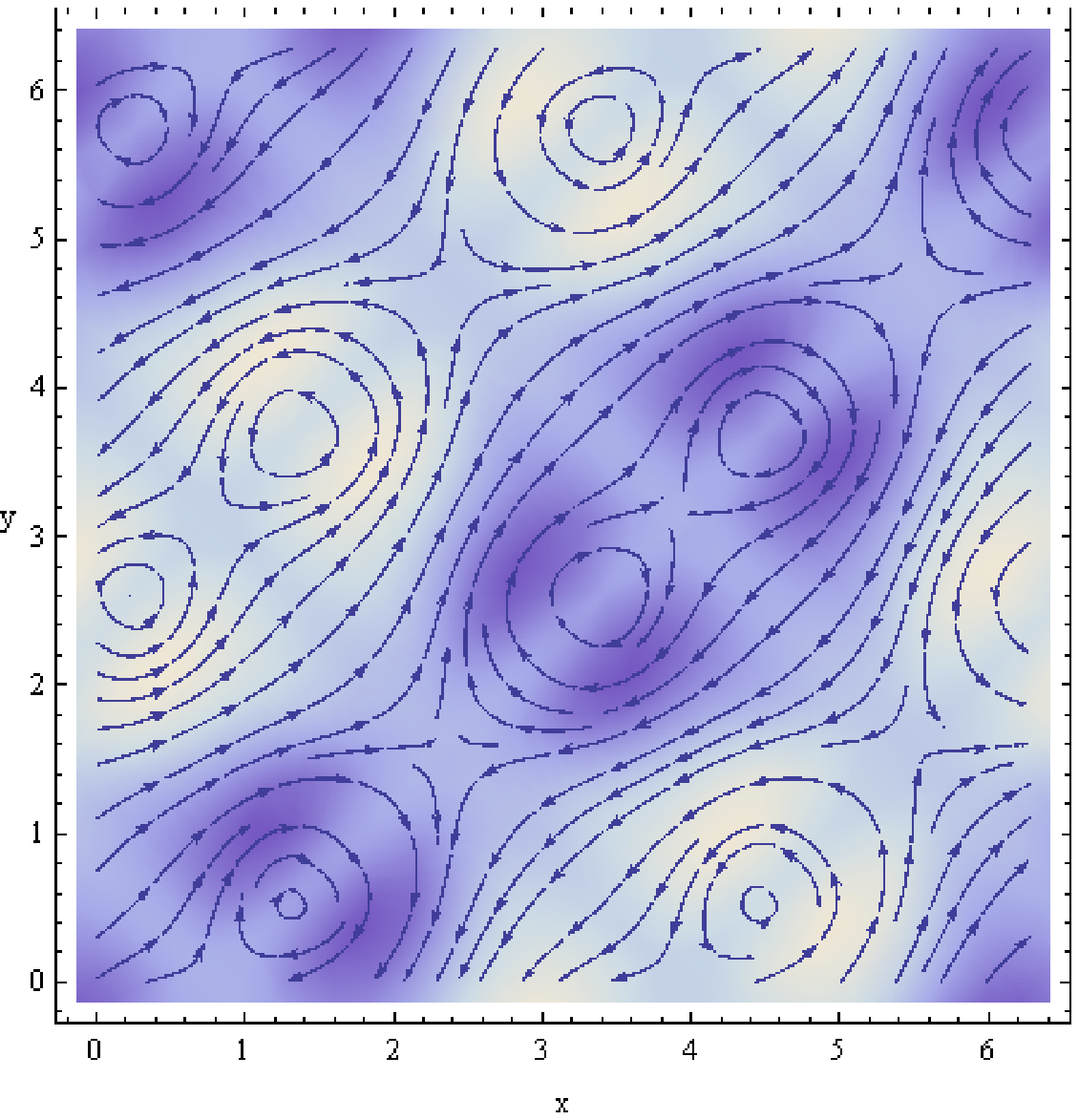}
    }
    \subfigure[]
    {
        \includegraphics[width=0.3\textwidth]{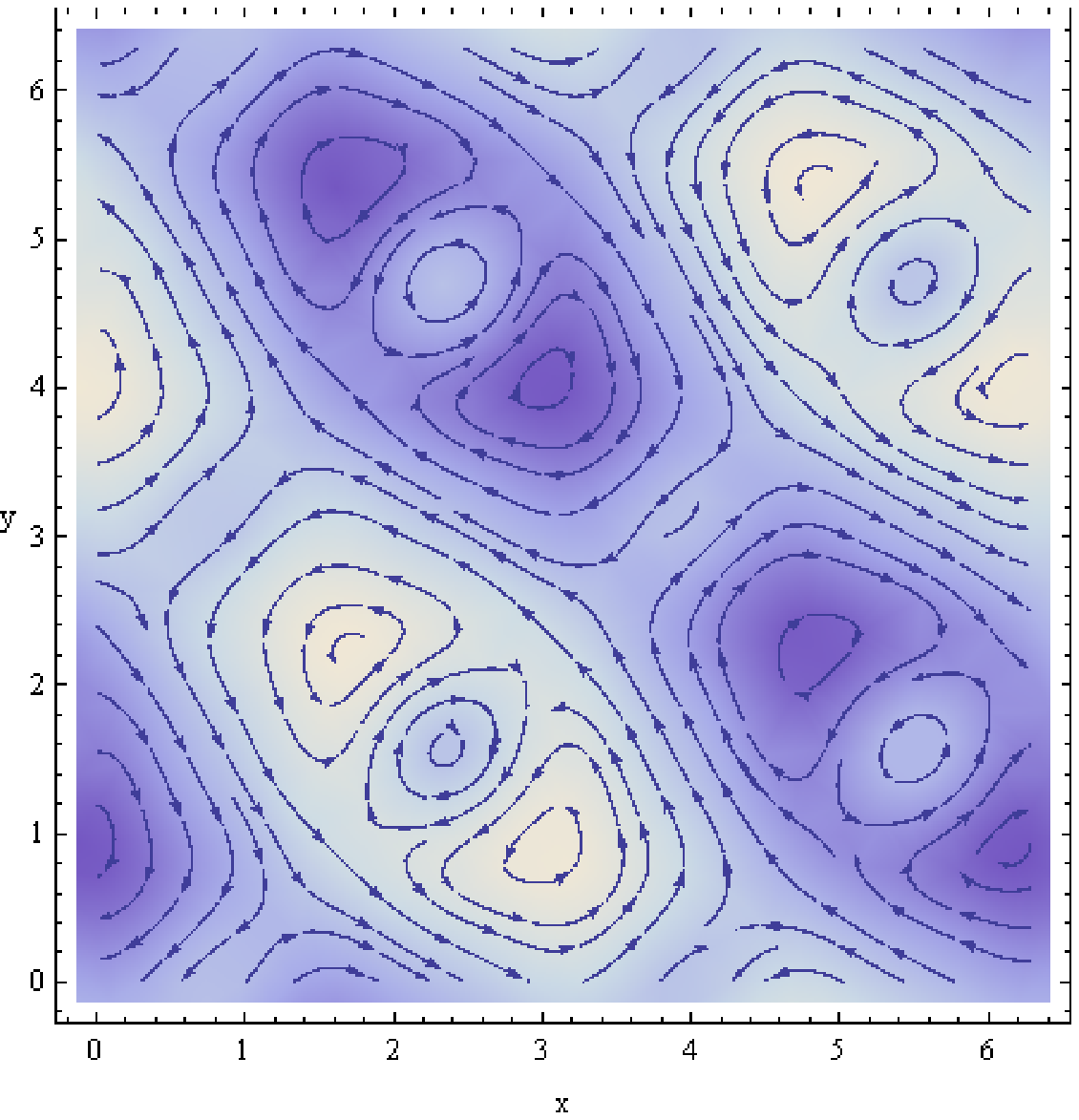}
    }
  \caption{(a) The equilibrium ABC field (\ref{eq:ABCnew}) with $B_1=B_2=1$, $B_3=0$, $\alpha=2>1$. The stream lines indicate the field components in the $x-y$ plane and color indicates the component perpendicular to the plane (same below for other vector fields).  (b) Trial perturbation $\vec{\xi}$ that renders negative potential energy (instability) for the equilibrium. We only plot $\vec{\xi}_{\perp}$, the components perpendicular to the equilibrium magnetic field. The perturbation is compressible in this case. (c) Perturbation magnetic field $\mathbf{B}_1$ resulted from the perturbation in (b).}\label{fig:Cartesian_compressible}
\end{figure}

\begin{figure}[htb]
  \centering
  \subfigure[]
    {
        \includegraphics[width=0.3\textwidth]{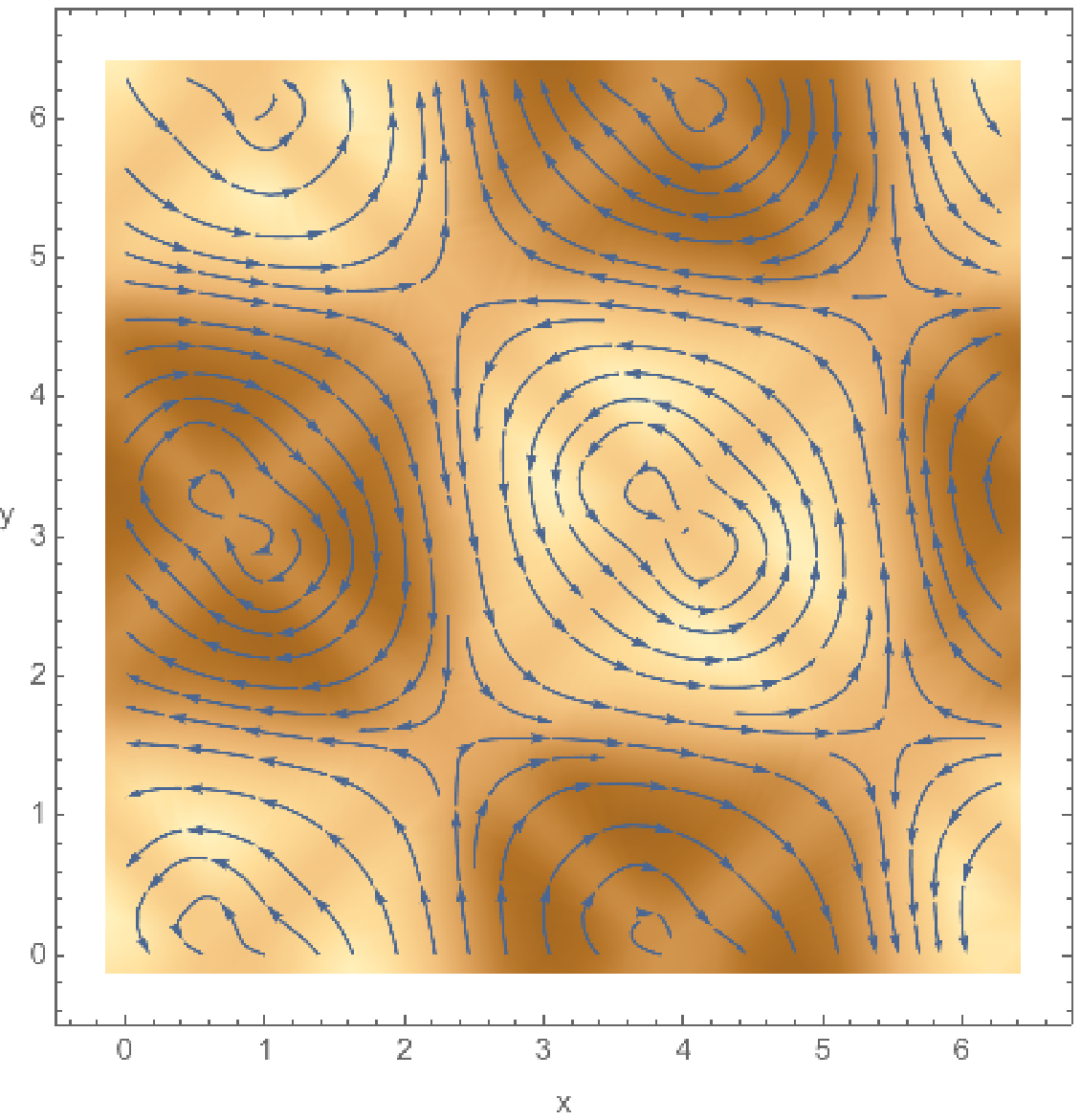}
    }
    \subfigure[]
    {
        \includegraphics[width=0.3\textwidth]{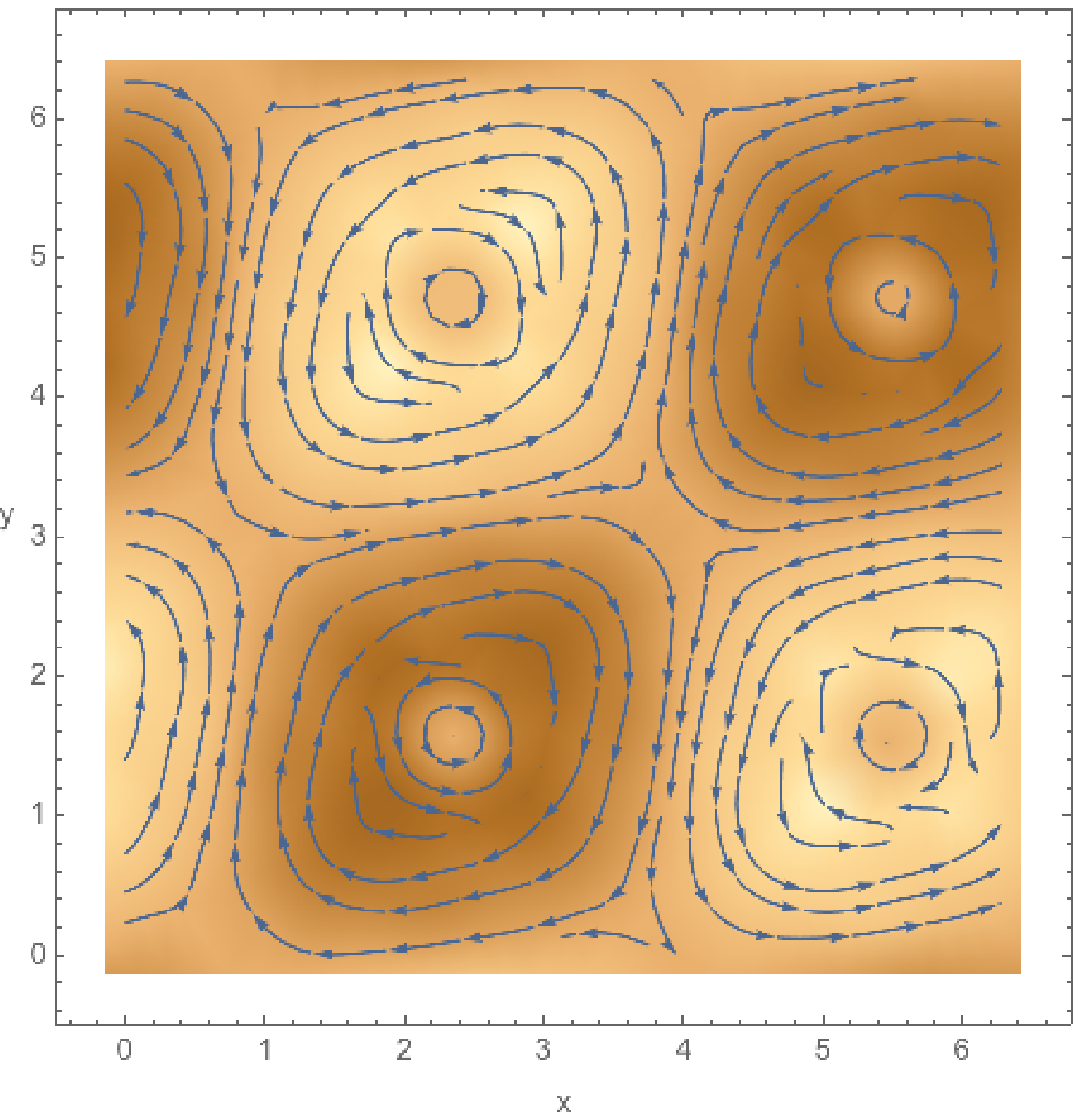}
    }
  \caption{(a) Similar to Figure \ref{fig:Cartesian_compressible} (b), this is the displacement field $\vec{\xi}_{\perp}$ derived from a trial function involving more Fourier modes. (b) Perturbation magnetic field $\mathbf{B}_1$ resulted from the perturbation in (a).}\label{fig:Cartesian_compressible_highk}
\end{figure}

The first case is 2D with $B_1=B_2=1$, $B_3=0$, $\alpha=2>1$.
We include several of the longest wavelength perturbations in the trial
function: $\mathbf{k}=(1,1,0)$, $(1,-1,0)$, $(1,3,0)$, $(1,-3,0)$, $(3,1,0)$,
$(3,-1,0)$, $(3,3,0)$, $(3,-3,0)$ plus their negative companions, then decide
the coefficients $\vec{\xi}_{\mathbf{k}}$ by minimizing the right hand side of
Equation (\ref{eq:variational}). The $\mathbf{k}$'s are chosen based on the
observation that in the numerical simulation, the dominant unstable mode is 2D
and has an electric field $\mathbf{E}\propto\vec{\xi}\times\mathbf{B}$ mainly
comprised of $\mathbf{k}=(\pm1,\pm1,0)$ components. We find that the
minimization does give negative $\omega^2$: $\omega^2<-0.04$, meaning a growth
rate of $\gamma>0.2$. As a comparison, the light crossing time scale $\tau=2\pi$
so $\gamma\tau>1.26$. This growth rate is less than what has been found from
numerical simulations ($\gamma\tau\approx5.5$ for this particular case) which is to be expected since we
can only get a lower limit on the growth rate from a variational approach. The Fourier components of this trial function are listed in Table \ref{table:ABC110}, and Figure \ref{fig:Cartesian_compressible} shows the stream plot of the perturbation. We find that including higher $\mathbf{k}$'s (but only $\mathbf{k}=(\pm(2s+1),\pm(2t+1),0), s,t\in\mathbb{Z}$ are relevant) allows us to get lower minimum $\omega^2$ (i.e. larger growth rate). For example, when the Fourier modes for the perturbation $\vec{\xi}$ include the following: $\mathbf{k}=(1,1,0)$, $(1,-1,0)$, $(1,3,0)$, $(1,-3,0)$, $(3,1,0)$, $(3,-1,0)$, $(3,3,0)$, $(3,-3,0)$, $(1,5,0)$, $(1,-5,0)$, $(5,1,0)$, $(5,-1,0)$, $(3,5,0)$, $(3,-5,0)$, $(5,3,0)$, $(5,-3,0)$, $(5,5,0)$, $(5,-5,0)$, we get the growth rate $\gamma>0.38$ and $\gamma\tau>2.38$. The corresponding trial function is plotted in Figure \ref{fig:Cartesian_compressible_highk}. Thus, the short wavelength perturbations are essential for the instability. Comparing these analytical trial functions with the dominant unstable mode discovered in numerical simulations, we find that the former is consistent with being truncated version of the latter.

\begin{table}[htdp]
\begin{center}
\begin{threeparttable}
\caption{Fourier components of the unstable trial function (compressible) for the equilibrium $B_1=B_2=1$, $B_3=0$, $\alpha=2$}\label{table:ABC110}
\begin{tabular}{ccc}
\hline
$\mathbf{k}_m$\tnote{$\dagger$} & $\Re\left(\vec{\xi} _m\right)$ & $\Im\left(\vec{\xi} _m\right)$\\
\hline
$\{1,1,0\}$ & $\{0.0795,0.0688,0.0641\}$ & $\{0.1367,0.0116,0.0502\}$\\
$\{1,-1,0\}$ & $\{-0.0623,-0.0569,0.1211\}$ & $\{-0.1611,-0.1557,0.0766\}$\\
$\{1,3,0\}$ & $\{-0.0646,0.0145,0.0474\}$ & $\{0.0342,0.0145,0.0515\}$\\
$\{1,-3,0\}$ & $\{-0.0400,0.0045,0.1013\}$ & $\{-0.0972,0.0045,-0.0441\}$\\
$\{3,1,0\}$ & $\{-0.0145,0.0700,-0.0420\}$ & $\{0.0145,0.0289,0.0568\}$\\
$\{3,-1,0\}$ & $\{0.0045,0.1083,0.0470\}$ & $\{-0.0045,-0.0511,0.1042\}$\\
$\{3,3,0\}$ & $\{0.0048,-0.0048,0.0008\}$ & $\{-0.0048,0.0048,0.0008\}$\\
$\{3,-3,0\}$ &  $\{-0.0154,-0.0154,0.0027\}$ & $\{0.0154,0.0154,0.0027\}$\\
\hline
\multicolumn{3}{c}{$\omega^2\le -0.04$, $\tau=2\pi$, $\omega\tau\ge1.26$}\\
\hline
\end{tabular}
\begin{tablenotes}
\item[$\dagger$] The coefficients for the $-\mathbf{k}_m$ terms are just the complex conjugate of the listed $\mathbf{k}_m$ coefficients.
\end{tablenotes}
\end{threeparttable}
\end{center}
\end{table}%

\begin{figure}[htb]
  \centering
  \subfigure[]
    {
        \includegraphics[width=0.3\textwidth]{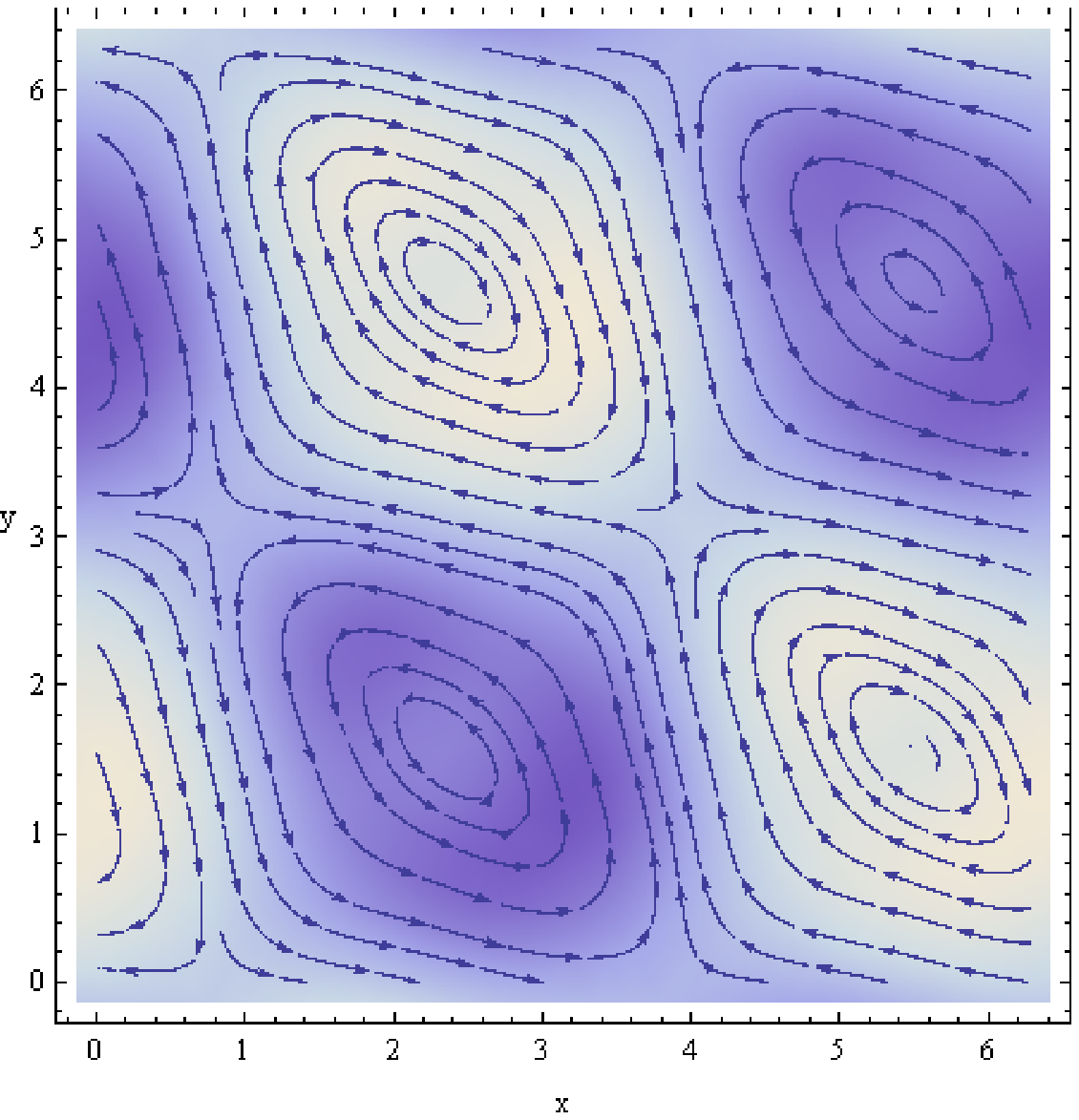}
    }
    \subfigure[]
    {
        \includegraphics[width=0.3\textwidth]{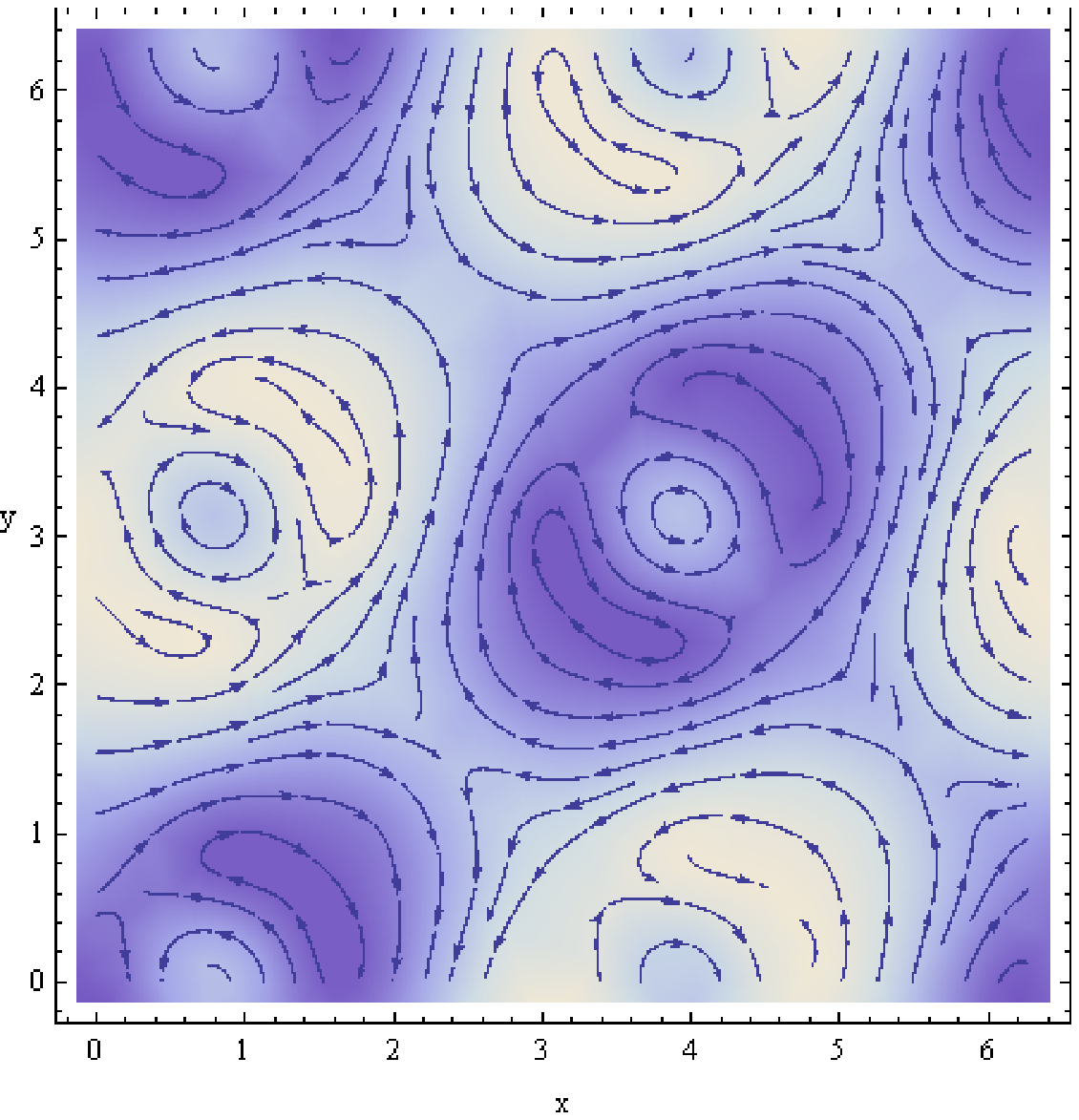}
    }
  \caption{(a) Incompressible trial perturbation $\vec{\xi}$ that renders negative potential energy (instability) for the equilibrium Figure \ref{fig:Cartesian_compressible}(a). (b) Perturbation magnetic field $\mathbf{B}_1$ resulted from the perturbation in (a).}\label{fig:Cartesian_incompressible}
\end{figure}

We also find that the instability still exist when incompressibility is imposed, i.e. $\nabla\cdot\vec{\xi}=0$. Table \ref{table:ABC110_incompressible} lists the Fourier components for the incompressible trial function that gives negative potential energy and Figure \ref{fig:Cartesian_incompressible} shows the corresponding stream plot.

\begin{table}[htdp]
\begin{center}
\begin{threeparttable}
\caption{Fourier components of the unstable trial function (incompressible) for the equilibrium $B_1=B_2=1$, $B_3=0$, $\alpha=2$}\label{table:ABC110_incompressible}
\begin{tabular}{ccc}
\hline
$\mathbf{k}_m$\tnote{$\dagger$} & $\Re\left(\vec{\xi} _m\right)$ & $\Im\left(\vec{\xi} _m\right)$\\
\hline
$\{1,1,0\}$ & $\{-0.1007,0.1007,0.0005\}$ & $\{-0.1007,0.1007,-0.0006\}$\\
$\{1,-1,0\}$ & $\{-0.0600,-0.0600,-0.0003\}$ & $\{-0.0600,-0.0600,0.0004\}$\\
$\{1,3,0\}$ & $\{-0.0108,0.0036,0.0001\}$ & $\{-0.0108,0.0036,0\}$\\
$\{1,-3,0\}$ & $\{-0.0181,-0.0060,-0.0001\}$ & $\{-0.0181,-0.0060,0.0001\}$\\
$\{3,1,0\}$ & $\{-0.0036,0.0108,-0.0001\}$ & $\{0.0036,-0.0108,0\}$\\
$\{3,-1,0\}$ & $\{-0.0060,-0.0181,0.0001\}$ & $\{0.0060,0.0181,0.0001\}$\\
$\{3,3,0\}$ & $\{-0.0139,0.0139,0\}$ & $\{0.0139,-0.0139,0\}$\\
$\{3,-3,0\}$ &  $\{-0.0083,-0.0083,0\}$ & $\{0.0083,0.0083,0\}$\\
\hline
\multicolumn{3}{c}{$\omega^2\le -0.016$, $\tau=2\pi$, $\omega\tau\ge0.795$}\\
\hline
\end{tabular}
\begin{tablenotes}
\item[$\dagger$] The coefficients for the $-\mathbf{k}_m$ terms are just the complex conjugate of the listed $\mathbf{k}_m$ coefficients.
\end{tablenotes}
\end{threeparttable}
\end{center}
\end{table}%

\begin{figure}[htb]
  \centering
  \subfigure[]
    {
        \includegraphics[width=0.3\textwidth]{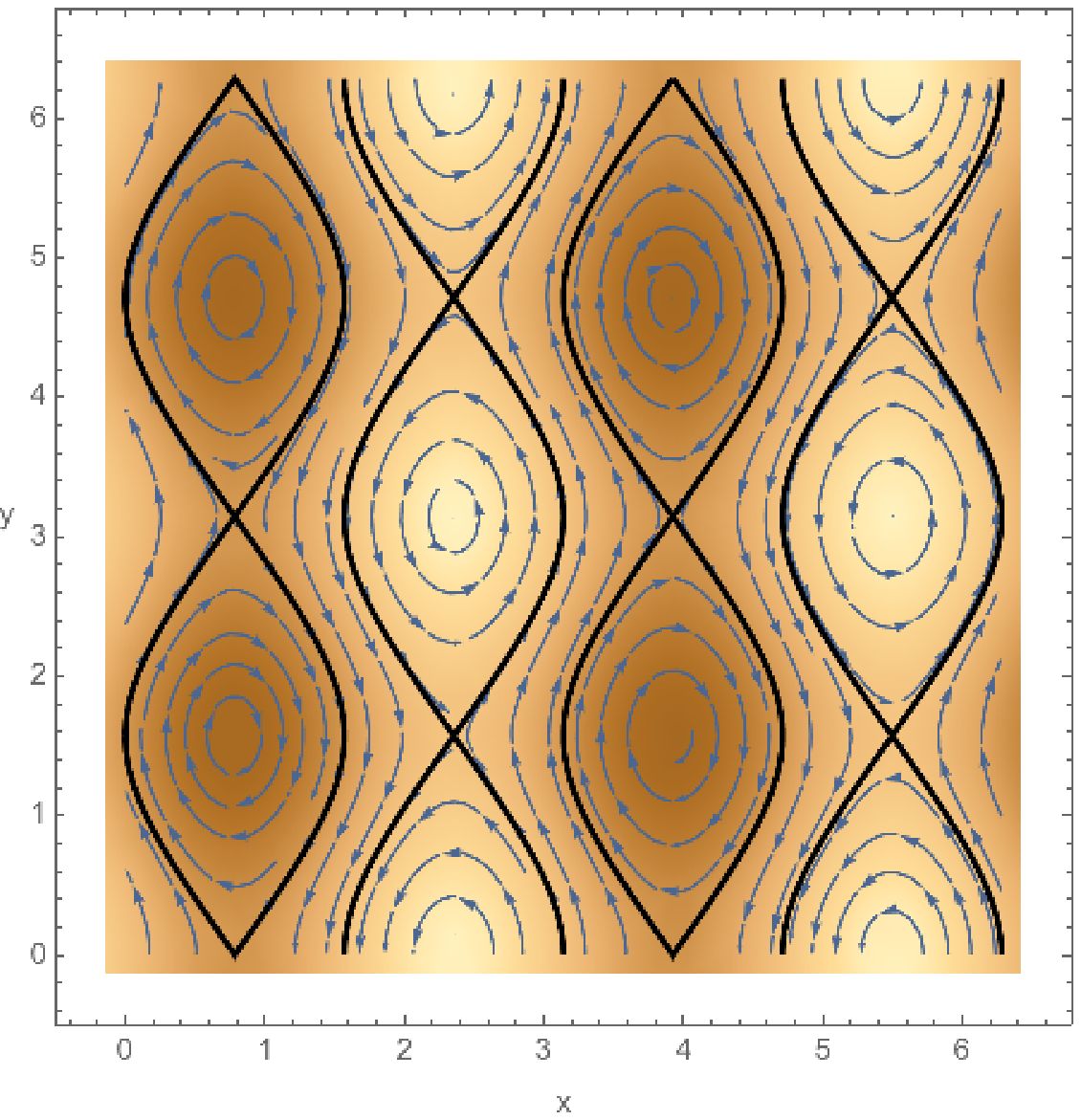}
    }
    \subfigure[]
    {
        \includegraphics[width=0.3\textwidth]{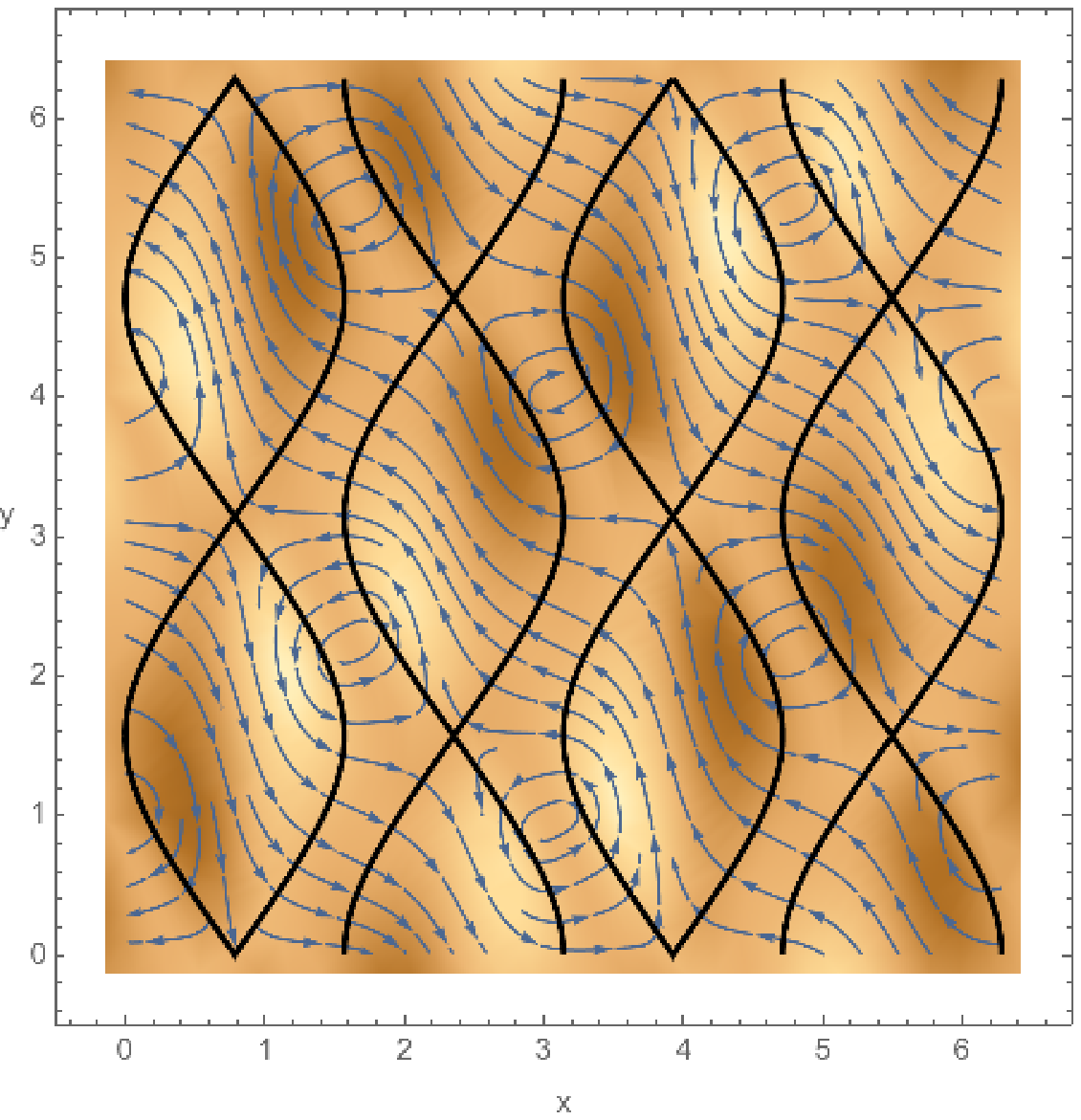}
    }
    \subfigure[]
    {
        \includegraphics[width=0.3\textwidth]{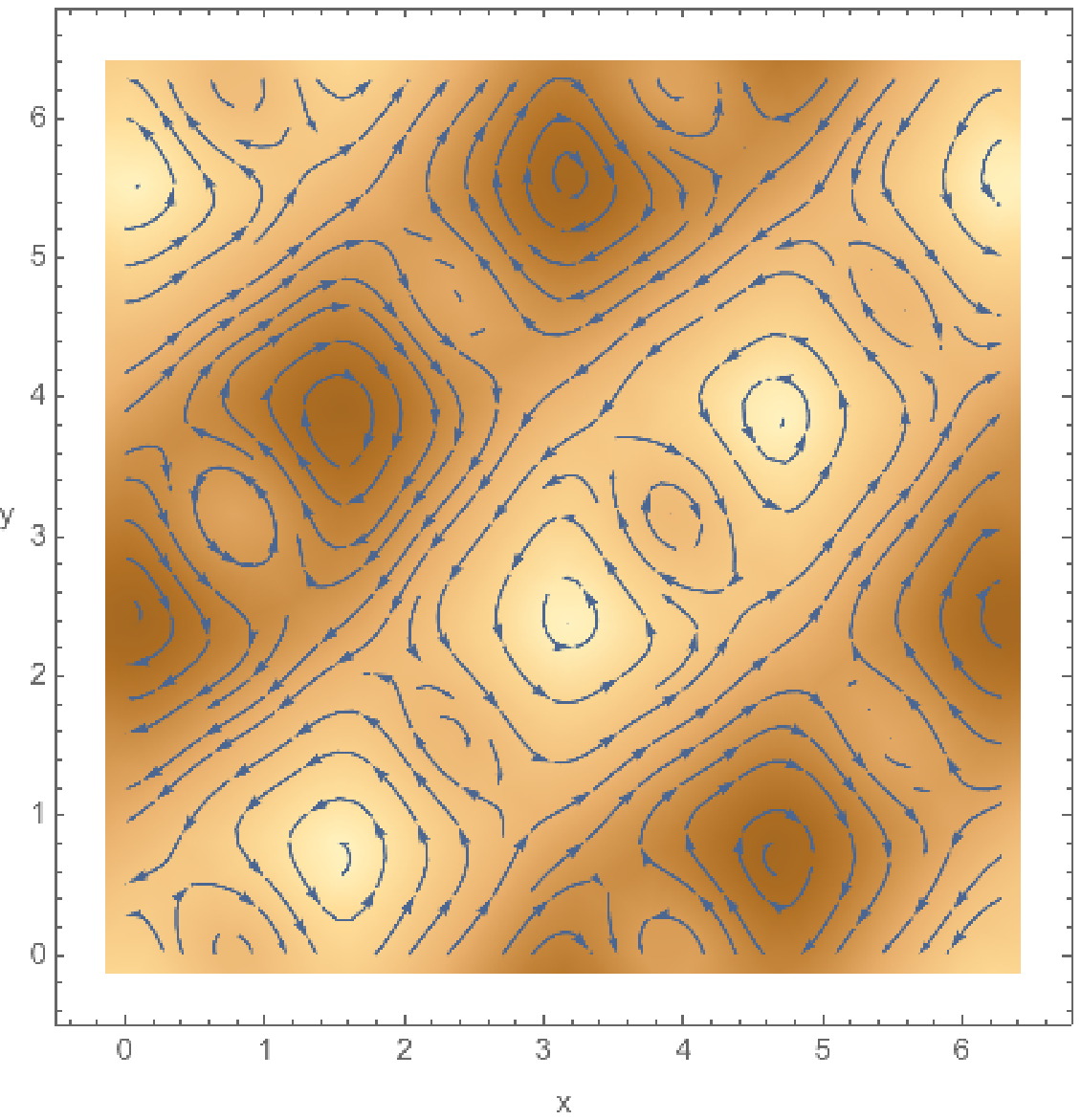}
    }
  \caption{(a) The equilibrium ABC field (\ref{eq:ABCnew}) with $B_1=1$, $B_2=1/2$, $B_3=0$, $\alpha=2>1$. The stream lines indicate the field components in the $x-y$ plane and color indicates the component perpendicular to the plane (same below for other vector fields). The thick black lines show the separatrices --- surfaces separating different topological domains --- in the equilibrium magnetic field. (b) Trial perturbation $\vec{\xi}$ that renders negative potential energy (instability) for the equilibrium. We only plot $\vec{\xi}_{\perp}$, the components perpendicular to the equilibrium magnetic field. The perturbation is compressible in this case. (c) Perturbation magnetic field $\mathbf{B}_1$ resulted from the perturbation in (b).}\label{fig:ABC105}
\end{figure}

Other more general 2D cases with $B_1\ne B_2$ and $B_1B_2\ne0$ are found to be unstable as well in numerical simulations, with growth rate decreasing from maximum to 0 as, say, $B_2/B_1$ goes from 1 to 0. As an illustration we apply our variational principle to the case $B_1=1$, $B_2=1/2$, $B_3=0$, using trial functions comprised of Fourier modes $\mathbf{k}=(1,1,0)$, $(1,-1,0)$, $(1,3,0)$, $(1,-3,0)$, $(3,1,0)$, $(3,-1,0)$, $(3,3,0)$, $(3,-3,0)$ and their negative companions. This is shown in Table \ref{table:ABC105} Figure \ref{fig:ABC105}, and the displacement field can be readily compared with Figure 3 in the main text. The growth rate we get from this trial function is $\gamma\tau>0.8$ --- the lower limit has been reduced from corresponding $B_1=B_2=1$ case.

\begin{table}[htdp]
\begin{center}
\begin{threeparttable}
\caption{Fourier components of the unstable trial function (compressible) for the equilibrium $B_1=1$,$B_2=1/2$, $B_3=0$, $\alpha=2$}\label{table:ABC105}
\begin{tabular}{ccc}
\hline
$\mathbf{k}_m$\tnote{$\dagger$} & $\Re\left(\vec{\xi} _m\right)$ & $\Im\left(\vec{\xi} _m\right)$\\
\hline
$\{1,1,0\}$ & $\{0.0914,-0.1849,0.0545\}$ & $\{0.1094,-0.2210,-0.0004\}$\\
$\{1,-1,0\}$ & $\{-0.0117,-0.0513,-0.1015\}$ & $\{0.0577,0.0876,-0.1069\}$\\
$\{1,3,0\}$ & $\{0.0306,-0.0022,-0.0378\}$ & $\{-0.0389,-0.0022,-0.0317\}$\\
$\{1,-3,0\}$ & $\{0.0274,0.0140,-0.0163\}$ & $\{0.0093,0.0140,0.0343\}$\\
$\{3,1,0\}$ & $\{0.0033,-0.0879,0.0619\}$ & $\{-0.0033,-0.0510,-0.0770\}$\\
$\{3,-1,0\}$ & $\{0.0209,0.0455,-0.0593\}$ & $\{-0.0209,-0.0094,-0.0233\}$\\
$\{3,3,0\}$ & $\{0.0166,-0.0184,0.0066\}$ & $\{-0.0166,0.0184,0.0066\}$\\
$\{3,-3,0\}$ &  $\{0.0026,0.0029,-0.0010\}$ & $\{-0.0026,-0.0029,-0.0010\}$\\
\hline
\multicolumn{3}{c}{$\omega^2\le -0.017$, $\tau=2\pi$, $\omega\tau\ge0.809$}\\
\hline
\end{tabular}
\begin{tablenotes}
\item[$\dagger$] The coefficients for the $-\mathbf{k}_m$ terms are just the complex conjugate of the listed $\mathbf{k}_m$ coefficients.
\end{tablenotes}
\end{threeparttable}
\end{center}
\end{table}%

\begin{figure}[htb]
  \centering
  \subfigure[]
    {
        \includegraphics[width=0.3\textwidth]{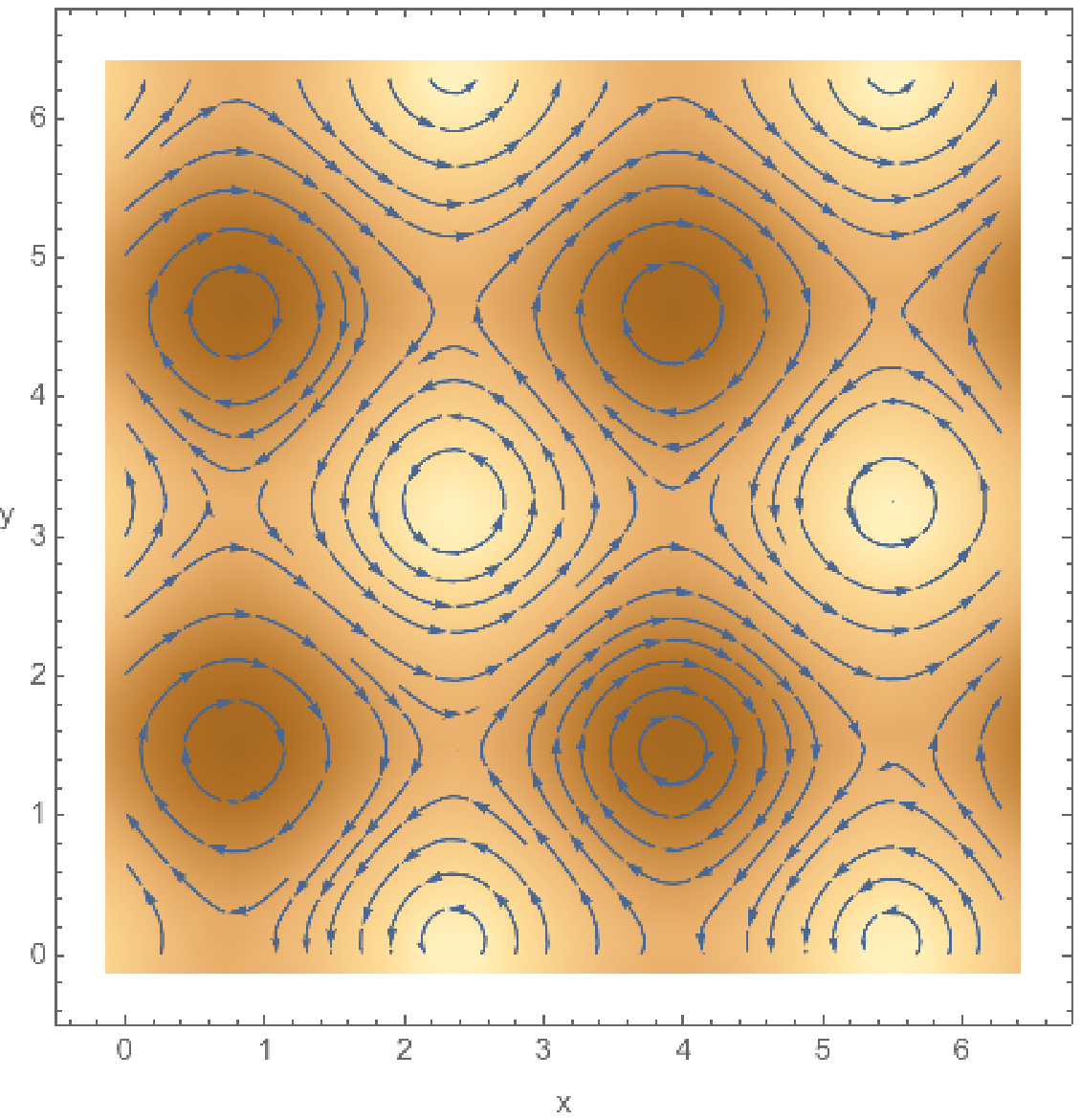}
    }
    \subfigure[]
    {
        \includegraphics[width=0.3\textwidth]{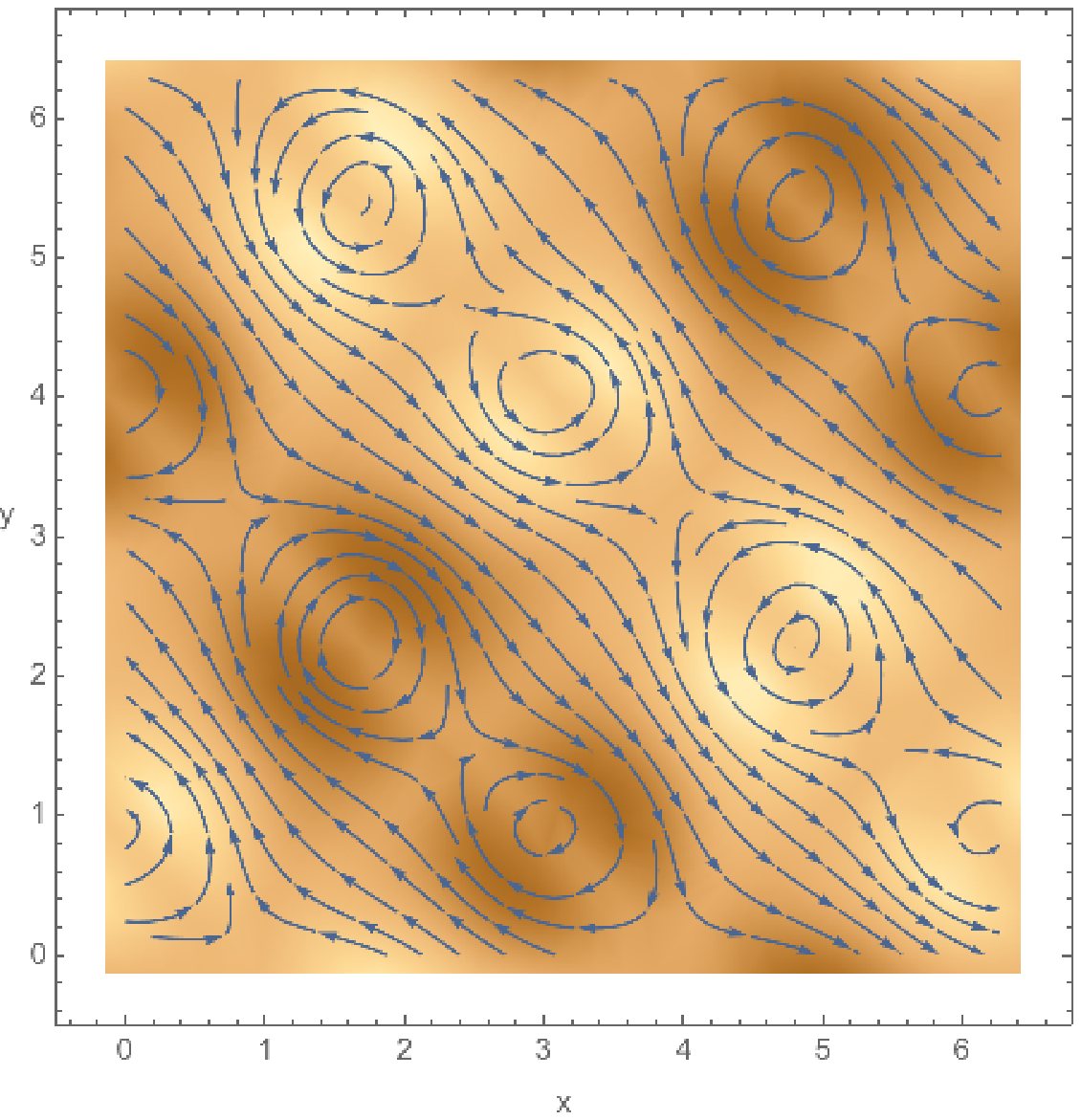}
    }
    \subfigure[]
    {
        \includegraphics[width=0.3\textwidth]{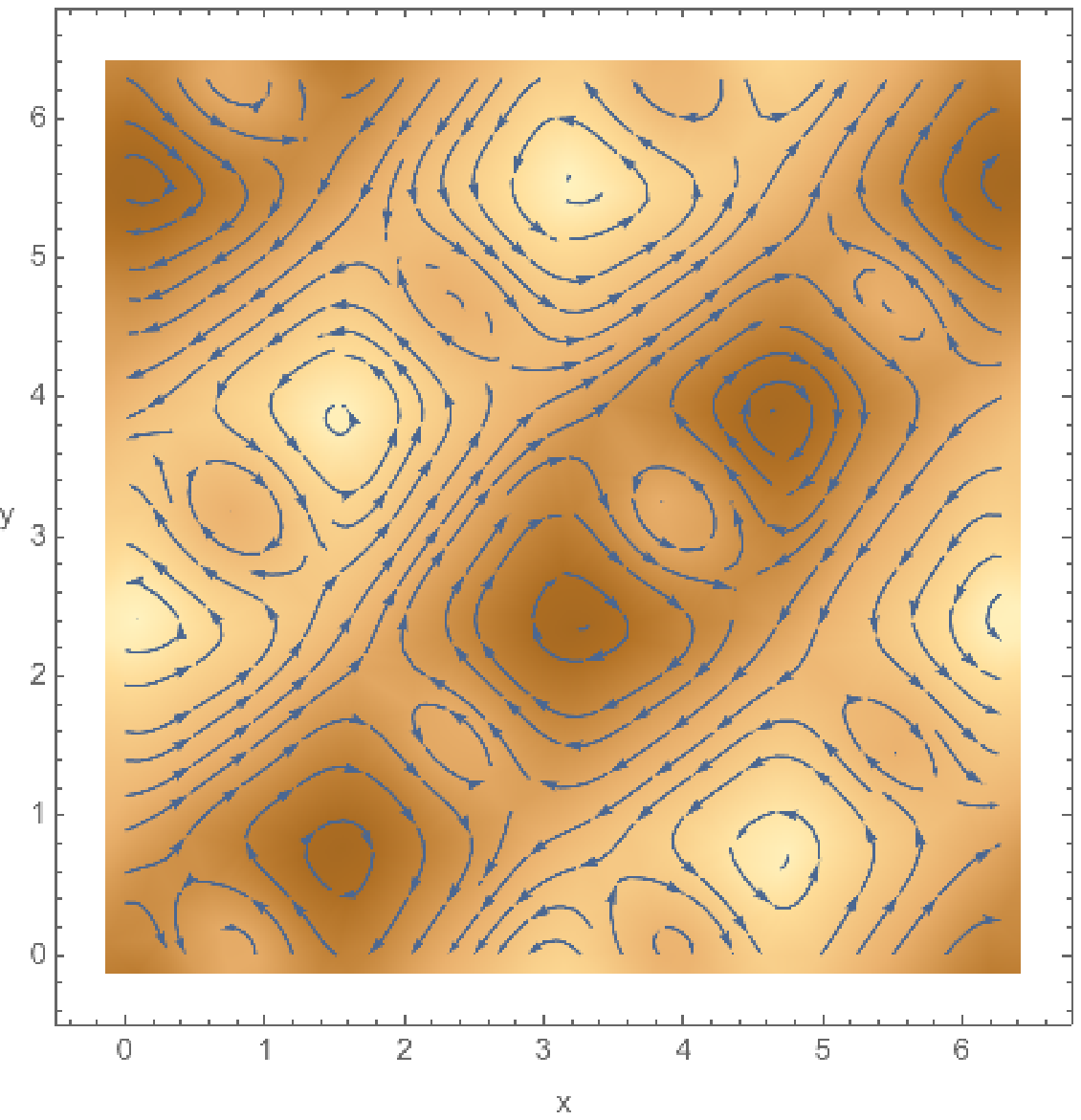}
    }
  \caption{(a) The equilibrium ABC field (\ref{eq:ABCnew}) with $B_1=B_2=1$, $B_3=1/5$, $\alpha=2>1$. The stream lines indicate the field components in the $z=0$ plane and color indicates the component perpendicular to the plane (same below for other vector fields). (b) Trial perturbation $\vec{\xi}$ that renders negative potential energy (instability) for the equilibrium. We only plot $\vec{\xi}_{\perp}$, the components perpendicular to the equilibrium magnetic field. The perturbation is compressible in this case. (c) Perturbation magnetic field $\mathbf{B}_1$ resulted from the perturbation in (b). Both (b) and (c) are stream plots on the $z=0$ plane.}\label{fig:ABC115}
\end{figure}

It is also instructive to consider a 3D example. Here we take $B_1=B_2=1$, $B_3=1/5$ and $\alpha=2$. Still use Fourier components $\mathbf{k}=(1,1,0)$, $(1,-1,0)$, $(1,3,0)$, $(1,-3,0)$, $(3,1,0)$, $(3,-1,0)$, $(3,3,0)$, $(3,-3,0)$ in the trial function, we found the unstable perturbation as shown in Table \ref{table:ABC115} and Figure \ref{fig:ABC115}.

\begin{table}[htdp]
\begin{center}
\begin{threeparttable}
\caption{Fourier components of the unstable trial function (compressible) for the equilibrium $B_1=B_2=1$, $B_3=1/5$, $\alpha=2$}\label{table:ABC115}
\begin{tabular}{ccc}
\hline
$\mathbf{k}_m$\tnote{$\dagger$} & $\Re\left(\vec{\xi} _m\right)$ & $\Im\left(\vec{\xi} _m\right)$\\
\hline
$\{1,1,0\}$ & $\{-0.1122,0.1121,-0.0231\}$ & $\{-0.1121,0.1120,0.0233\}$\\
$\{1,-1,0\}$ & $\{-0.0180,-0.0180,0.0039\}$ & $\{-0.0180,-0.0180,-0.0036\}$\\
$\{1,3,0\}$ & $\{-0.0029,0.0024,-0.0007\}$ & $\{-0.0029,0.0024,0.0008\}$\\
$\{1,-3,0\}$ & $\{-0.0182,-0.0147,0.0048\}$ & $\{-0.0182,-0.0147,-0.0047\}$\\
$\{3,1,0\}$ & $\{-0.0024,0.0029,0.0007\}$ & $\{0.0024,-0.0029,0.0008\}$\\
$\{3,-1,0\}$ & $\{-0.0147,-0.0182,-0.0048\}$ & $\{0.0147,0.0183,-0.0047\}$\\
$\{3,3,0\}$ & $\{-0.0151,0.0151,-0.0026\}$ & $\{0.0151,-0.0151,-0.0026\}$\\
$\{3,-3,0\}$ &  $\{-0.0024,-0.0024,0.0004\}$ & $\{0.0024,0.0024,0.0004\}$\\
\hline
\multicolumn{3}{c}{$\omega^2\le -0.016$, $\tau=2\pi$, $\omega\tau\ge0.801$}\\
\hline
\end{tabular}
\begin{tablenotes}
\item[$\dagger$] The coefficients for the $-\mathbf{k}_m$ terms are just the complex conjugate of the listed $\mathbf{k}_m$ coefficients.
\end{tablenotes}
\end{threeparttable}
\end{center}
\end{table}%

\bibliographystyle{h-physrev}
\bibliography{library,ref,FFref}

\end{document}